\documentclass[twocolumn,twocolappendix]{aastex631}

\usepackage{amsmath}
\usepackage{longtable}
\usepackage{graphicx}

\usepackage{soul}

\begin{document}

\title{exoALMA XI: ALMA Observations and Hydrodynamic Models of LkCa~15: Implications for Planetary Mass Companions in the Dust Continuum Cavity    }

\author[0009-0003-8984-2094]{Charles H. Gardner}
\affiliation{Department of Physics and Astronomy, Rice University,
Houston, TX 77005, USA}
\affiliation{Los Alamos National Laboratory, Los Alamos, NM 87545, USA}

\author[0000-0001-8061-2207]{Andrea Isella}
\affiliation{Department of Physics and Astronomy, Rice University,
Houston, TX 77005, USA}
\affiliation{Rice Space Insitute, Rice University, Houston, TX 77005, USA}

\author[0000-0003-3556-6568]{Hui Li}
\affiliation{Los Alamos National Laboratory, Los Alamos, NM 87545, USA}

\author[0000-0002-4142-3080]{Shengtai Li}
\affiliation{Los Alamos National Laboratory, Los Alamos, NM 87545, USA}

\author[0000-0001-7258-770X]{Jaehan Bae}\affiliation{Department of Astronomy, University of Florida, Gainesville, FL 32611, USA}

\author[0000-0001-6378-7873]{Marcelo Barraza-Alfaro}\affiliation{Department of Earth, Atmospheric, and Planetary Sciences, Massachusetts Institute of Technology, Cambridge, MA 02139, USA}

\author[0000-0002-7695-7605]{Myriam Benisty}\affiliation{Université Côte d'Azur, Observatoire de la Côte d'Azur, CNRS, Laboratoire Lagrange, France}\affiliation{Max-Planck Institute for Astronomy (MPIA), Königstuhl 17, 69117 Heidelberg, Germany}

\author[0000-0002-2700-9676]{Gianni Cataldi}
\affiliation{National Astronomical Observatory of Japan, Osawa 2-21-1, Mitaka, Tokyo 181-8588, Japan}

\author[0000-0003-2045-2154]{Pietro Curone}\affiliation{Dipartimento di Fisica, Universit\`a degli Studi di Milano, Via Celoria 16, 20133 Milano, Italy}\affiliation{Departamento de Astronomía, Universidad de Chile, Camino El Observatorio 1515, Las Condes, Santiago, Chile}

\author[0000-0002-1031-4199]{Josh A. Eisner}\affiliation{Steward Observatory, University of Arizona, 933 N. Cherry Ave., Tucson, AZ 85719, USA}

\author[0000-0003-4689-2684]{Stefano Facchini}
\affiliation{Dipartimento di Fisica, Universit\`a degli Studi di Milano, Via Celoria 16, 20133 Milano, Italy}

\author[0000-0003-4679-4072]{Daniele Fasano}\affiliation{Université Côte d'Azur, Observatoire de la Côte d'Azur, CNRS, Laboratoire Lagrange, France}

\author[0000-0002-9298-3029]{Mario Flock} 
\affiliation{Max-Planck Institute for Astronomy (MPIA), Königstuhl 17, 69117 Heidelberg, Germany}

\author[0000-0002-7821-0695]{Katherine B. Follette}\affiliation{Amherst College Dept. Of Physics and Astronomy, 25 East Dr, Amherst MA 01002 USA}

\author[0000-0003-1117-9213]{Misato Fukagawa} 
\affiliation{National Astronomical Observatory of Japan, 2-21-1 Osawa, Mitaka, Tokyo 181-8588, Japan}

\author[0000-0002-5503-5476]{Maria Galloway-Sprietsma}\affiliation{Department of Astronomy, University of Florida, Gainesville, FL 32611, USA}

\author[0000-0002-5910-4598]{Himanshi Garg}
\affiliation{School of Physics and Astronomy, Monash University, Clayton VIC 3800, Australia}

\author[0000-0002-8138-0425]{Cassandra Hall}\affiliation{Department of Physics and Astronomy, The University of Georgia, Athens, GA 30602, USA}\affiliation{Center for Simulational Physics, The University of Georgia, Athens, GA 30602, USA}\affiliation{Institute for Artificial Intelligence, The University of Georgia, Athens, GA, 30602, USA}

\author[0000-0001-6947-6072]{Jane Huang}\affiliation{Department of Astronomy, Columbia University, 538 W. 120th Street, Pupin Hall, New York, NY 10027, United States of America}

\author[0000-0003-1008-1142]{John~D.~Ilee} 
\affiliation{School of Physics and Astronomy, University of Leeds, Leeds, UK, LS2 9JT}

\author[0000-0002-6194-043X]{Michael J. Ireland}\affiliation{Research School of Astronomy and Astrophysics, Australian National University, Canberra 2611, Australia}

\author[0000-0001-8446-3026]{Andr\'es F. Izquierdo} 
\affiliation{Department of Astronomy, University of Florida, Gainesville, FL 32611, USA}
\affiliation{Leiden Observatory, Leiden University, P.O. Box 9513, NL-2300 RA Leiden, The Netherlands}
\affiliation{European Southern Observatory, Karl-Schwarzschild-Str. 2, D-85748 Garching bei M\"unchen, Germany}

\author[0000-0002-8828-6386]{Christopher M. Johns-Krull}\affiliation{Department of Physics and Astronomy, Rice University,
Houston, TX 77005, USA}

\author[0000-0001-7235-2417]{Kazuhiro Kanagawa} 
\affiliation{College of Science, Ibaraki University, 2-1-1 Bunkyo, Mito, Ibaraki 310-8512, Japan}

\author[0000-0001-9811-568X]{Adam L. Kraus}\affiliation{Department of Astronomy, University of Texas at Austin, 2515 Speedway Stop C1400, Austin, TX 78712, USA}

\author[0000-0002-8896-9435]{Geoffroy Lesur} 
\affiliation{Univ. Grenoble Alpes, CNRS, IPAG, 38000 Grenoble, France}

\author[0000-0002-9442-137X]{Shangfei Liu} \affiliation{School of Physics and Astronomy, Sun Yat-sen University, Zhuhai 519082, People's Republic of China}\affiliation{CSST Science Center for the Guangdong-HongKong-Macau Great Bay Area, Sun Yat-sen University, Zhuhai 519082, People's Republic of China}

\author[0000-0003-4663-0318]{Cristiano Longarini}\affiliation{Institute of Astronomy, University of Cambridge, Madingley Rd, CB30HA, Cambridge, UK}\affiliation{Dipartimento di Fisica, Universit\`a degli Studi di Milano, Via Celoria 16, 20133 Milano, Italy}

\author[0000-0002-8932-1219]{Ryan A. Loomis}
\affiliation{National Radio Astronomy Observatory, 520 Edgemont Rd., Charlottesville, VA 22903, USA}


\author[0000-0002-1637-7393]{Francois Menard}\affiliation{Univ. Grenoble Alpes, CNRS, IPAG, 38000 Grenoble, France}

\author[0000-0003-4039-8933]{Ryuta Orihara} 
\affiliation{College of Science, Ibaraki University, 2-1-1 Bunkyo, Mito, Ibaraki 310-8512, Japan}

\author[0000-0001-5907-5179]{Christophe Pinte}\affiliation{Univ. Grenoble Alpes, CNRS, IPAG, 38000 Grenoble, France}\affiliation{School of Physics and Astronomy, Monash University, VIC 3800, Australia}

\author[0000-0002-4716-4235]{Daniel Price}\affiliation{School of Physics and Astronomy, Monash University, VIC 3800, Australia}

\author[0000-0001-8123-2943]{Luca Ricci}\affiliation{Department of Physics and Astronomy, California State University Northridge, 18111 Nordhoff Street, Northridge, CA 91130, USA}

\author[0000-0003-4853-5736]{Giovanni Rosotti}\affiliation{Dipartimento di Fisica, Universit\`a degli Studi di Milano, Via Celoria 16, 20133 Milano, Italy}

\author[0000-0001-6871-6775]{Steph Sallum} \affiliation{Department of Physics and Astronomy, University of California, Irvine, 4129 Frederick Reines Hall, Irvine, CA, USA}

\author[0000-0002-0491-143X]{Jochen Stadler}\affiliation{Universit\'{e} C\^{o}te d'Azur, Observatoire de la C\^{o}te d'Azur, CNRS, Laboratoire Lagrange, France}\affiliation{Univ. Grenoble Alpes, CNRS, IPAG, 38000 Grenoble, France}

\author[0000-0003-1534-5186]{Richard Teague}\affiliation{Department of Earth, Atmospheric, and Planetary Sciences, Massachusetts Institute of Technology, Cambridge, MA 02139, USA}

\author[0000-0002-3468-9577]{Gaylor Wafflard-Fernandez} 
\affiliation{Univ. Grenoble Alpes, CNRS, IPAG, 38000 Grenoble, France}

\author[0000-0003-1526-7587]{David J. Wilner}\affiliation{Center for Astrophysics | Harvard \& Smithsonian, Cambridge, MA 02138, USA}

\author[0000-0002-7501-9801]{Andrew J. Winter}\affiliation{Universit\'{e} C\^{o}te d'Azur, Observatoire de la C\^{o}te d'Azur, CNRS, Laboratoire Lagrange, 06300 Nice, France}\affiliation{Max-Planck Institute for Astronomy (MPIA), Königstuhl 17, 69117 Heidelberg, Germany}

\author[0000-0002-7212-2416]{Lisa W\"olfer} 
\affiliation{Department of Earth, Atmospheric, and Planetary Sciences, Massachusetts Institute of Technology, Cambridge, MA 02139, USA}

\author[0000-0003-1412-893X]{Hsi-Wei Yen} 
\affiliation{Academia Sinica Institute of Astronomy \& Astrophysics, 11F of Astronomy-Mathematics Building, AS/NTU, No.1, Sec. 4, Roosevelt Rd, Taipei 10617, Taiwan}

\author[0000-0001-8002-8473]{Tomohiro C. Yoshida}\affiliation{National Astronomical Observatory of Japan, 2-21-1 Osawa, Mitaka, Tokyo 181-8588, Japan}\affiliation{Department of Astronomical Science, The Graduate University for Advanced Studies, SOKENDAI, 2-21-1 Osawa, Mitaka, Tokyo 181-8588, Japan}

\author[0000-0001-9319-1296]{Brianna Zawadzki} 
\affiliation{Department of Astronomy, Van Vleck Observatory, Wesleyan University, 96 Foss Hill Drive, Middletown, CT 06459, USA}
\affiliation{Department of Astronomy \& Astrophysics, 525 Davey Laboratory, The Pennsylvania State University, University Park, PA 16802, USA}

\author[0000-0003-3616-6822]{Zhaohuan Zhu}\affiliation{Department of Physics and Astronomy, University of Nevada, Las Vegas, 4505 S. Maryland Pkwy, Las Vegas, NV, 89154, USA}\affiliation{Nevada Center for Astrophysics, University of Nevada, Las Vegas, 4505 S. Maryland Pkwy, Las Vegas, NV, 89154, USA}



\begin{abstract}

In the past decade, the Atacama Large Millimeter/submillimeter Array (ALMA) has revealed a plethora of substructures in the disks surrounding young stars. 
These substructures have several proposed formation mechanisms, with one leading theory being the interaction between the disk and newly formed planets. 
In this Letter, we present high angular resolution ALMA observations of LkCa~15's disk that reveal a striking difference in dust and CO emission morphology. 
The dust continuum emission shows a ring-like structure characterized by a dust-depleted inner region of $\sim$40 au in radius. 
Conversely, the CO emission is radially smoother and shows no sign of gas depletion within the dust cavity.  
We compare the observations with models for the disk-planet interaction, including radiative transfer calculation in the dust and CO emission.  
This source is particularly interesting as the presence of massive planets within the dust cavity has been suggested based on previous NIR observations. 
We find that the level of CO emission observed within the dust cavity is inconsistent with the presence of planets more massive than Jupiter orbiting between 10-40 au.  
Instead, we argue that the LkCa~15 innermost dust cavity might be created either by a chain of low-mass planets, or by other processes that do not require the presence of planets.

\end{abstract}



\section{Introduction} \label{sec:intro}

In the age of the Atacama Large Millimeter/submillimeter Array (ALMA), dust continuum observations of the protoplanetary disks (PPDs) around young stars have revealed a wide variety of substructures such as rings, gaps, spirals, and crescents \citep[see, e.g., ][]{Andrews+2018}. 
Many theories exist about the origin of these substructures \citep{Bae+2023}. 
Yet, the most widely accepted is that they arise due to the interaction between newly formed planets and the circumstellar disk. 
This hypothesis was confirmed in the case of PDS~70 with the direct detection of two planets orbiting inside a dust- and gas-depleted cavity imaged with ALMA \citep{Keppler+2018, Keppler+2019, Muller+2018, Wagner+2018,Christiaens+2019a,Christiaens+2019b,Christiaens+2024,Haffert+2019}\citep{Stolker+2020,Facchini+2021,Wang+2021,Zhou+2021,Perotti+2023,Blakely+2024,Close+2025}. 
Additionally, ALMA observations revealed a circumplanetary disk (CPD) around PDS~70~c, confirming the detection at optical wavelengths and providing a new tool to discover and characterize planets at mm-wavelengths \citep{Isella+2019, Benisty+2021}. 
Yet, despite the large number of substructures detected in the dust continuum in various circumstellar disks, the number of (candidate) planets in a PPD remains low, as many of these planets are thought to be just at or below current detection capabilities \citep{Asensio-Torres+2021}. 
Therefore, simultaneous observation and theoretical modeling can work in tandem to constrain planet masses and locations in sources where distinct substructures are visible, yet detections remain elusive \citep{Toci+2020}. The LkCa~15 disk provides an outstanding example of one such source.

LkCa~15 is a 1.25 $M_{\odot} \pm 0.1 M_{\odot}$ star located in the Taurus star-forming region \citep{Donati+2019} at a distance of $\approx$157.2 pc \citep{Gaia+2023}. 
The star is surrounded by a disk characterized by a 90au-wide cavity and three dusty rings seen in the mm-wave continuum emission \citep{Pietu+2006, Andrews+2011, Isella+2012, Jin+2019, Facchini+2020}, which have been hypothesized to be the product of gravitational perturbations by planets. 
In particular, the large dust cavity may have been cleared by one or more planets, while the dust rings have been shown to be consistent with perturbations from planets with masses between 0.1-0.3 Jupiter masses ($M_J$) \citep{Leemker+2022,Facchini+2020}.  
This hypothesis found further support in recent ALMA observations that revealed clumps along the innermost dust ring consistent with dust trapped at the $L_{4}$ and $L_{5}$ Lagrangian points of a 0.1-0.3 $M_J$ planet orbiting at 42 au from the central star \citep{Long+2022}.

However, the more-than-a-decade-long direct search for planets in the LkCa~15 disk has not yet yielded conclusive results.  
Using Keck NIR sparse aperture masking observations (SAM), \citet{Kraus+2012} claimed a likely detection of a $\sim 6$ $M_J$ companion, LkCa~15~b, within the dust central cavity at an orbital radius of 15 au. 
This discovery was seemingly confirmed by \citet{Sallum+2015}, who claimed the detection of LkCa~15~b in NIR SAM images acquired with the Large Binocular Telescope and adaptive optics high-contrast images of the H$\alpha$ emission line obtained with the Magellan telescope. 
The latter suggested that the planet was accreting circumstellar material. 
Furthermore, they reported detecting an additional NIR compact source inside the mm-wave dust cavity attributed to a second planet, LkCa~15~c, orbiting at 19 au from the central star. 
By comparing the observed fluxes with models including the emission from both the planet and a CPD, they concluded that LkCa~15~b and c have masses between 1-5 $M_J$. 
However, the presence of massive planets orbiting within LkCa~15's disk dust cavity has been disputed by more recent studies that revealed scattered light emission arising from this region, and, more importantly, at the presumed location of LkCa~15~b and c \citep{Oh+2016, Thalmann+2016, Currie+2019}. 
These studies suggested that spatial filtering by SAM of the scattered light from the disk inner rim, including scattered H$\alpha$ stellar emission, could have been misinterpreted as planetary emission by previous studies \citep[see, also,][]{Blakely+2022}.

To shed light on the origin of the complex morphology of LkCa~15's disk, and, in particular, on the presence of planets within the mm-wave dust cavity, we analyze in this work ALMA observations that map the 0.87~mm dust continuum and $^{12}$CO J=3-2 line emission at an unprecedented spatial resolution of $\sim$6 au in both tracers, resolving the dust cavity across more than 15 resolution elements.  
Until recently, studies on the presence of planets in circumstellar disks based on ALMA observations have focused on maps of the dust continuum emission \citep[see, e.g.,][]{Zhang+2018}.   
Continuum observations alone, however, provide only partial information on the distribution of the circumstellar material. 
In particular, planet mass estimates based on the width of dust cavities and gaps require assumptions on the dust-gas coupling, which, in turn, depends on poorly known quantities such as the gas density and kinematics, as well as the dust grain size and composition.
Thus, observations of molecular gas at the same angular scales of the continuum are essential to constrain the distribution of the circumstellar material and test the planet-disk interaction hypothesis. However, since molecular line observations that match the angular resolution and sensitivity of the continuum require much longer observations, they are available for very few objects \citep{vandermarel+2016,Fedele+2017,Isella+2018,Gabellini+2019}. 

This Letter is structured as follows. In Section \ref{sec:data},  we describe the data acquisition and imaging procedure. Section \ref{sec:obs} presents and describes our continuum and CO observations. Section \ref{sec:mod} describes our hydrodynamic and radiative transfer modeling. Finally, in Section \ref{sec:discuss}, we discuss our results and their implications.


\section{Data} \label{sec:data}

\begin{figure*}[!th]
    \centering
    \includegraphics[width=\linewidth]{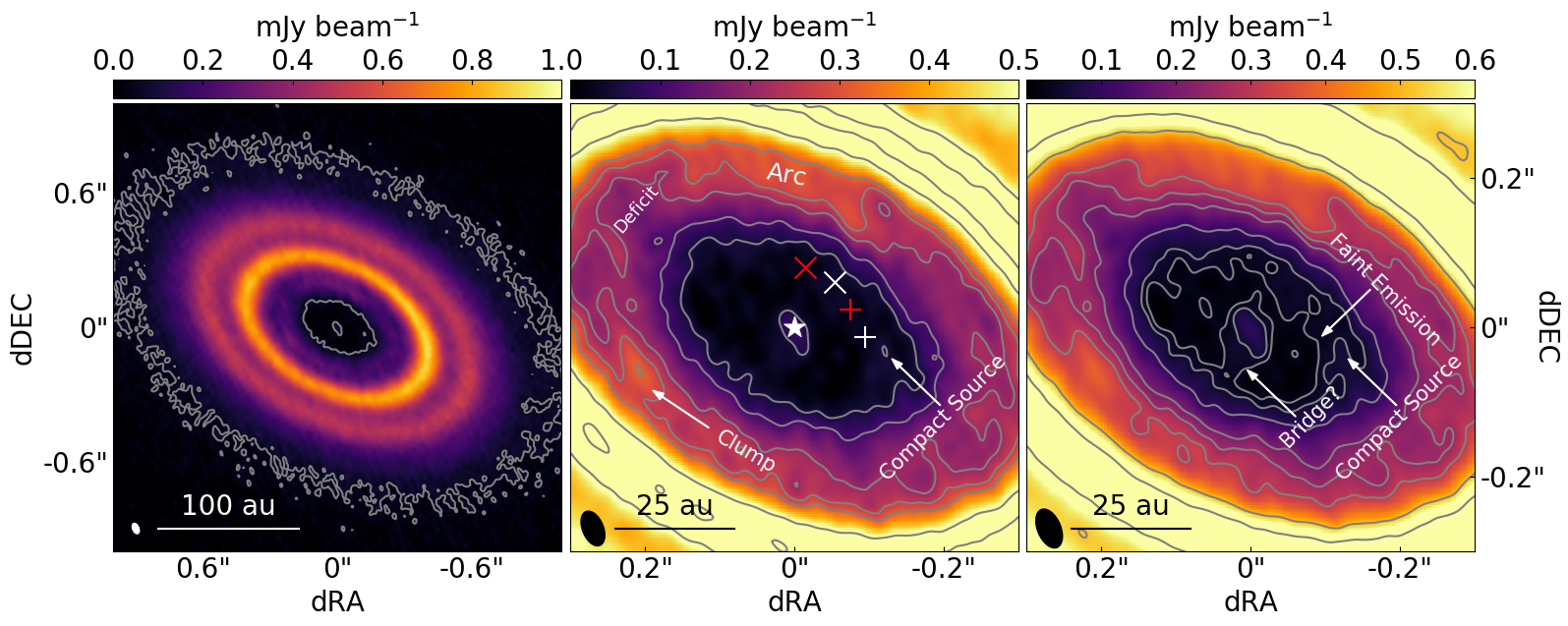}
    \caption{Left: The 0.87~mm continuum image of LkCa~15's disk obtained with \textit{robust}$=0.7$ corresponding to a beam FWHM of $0.049\arcsec\times0.028\arcsec$. The disk exhibits a three-ringed morphology with a faint inner ring, a bright middle ring, and an intermediate brightness outer ring. A continuum contour at 5 times the rms noise of 11 $\mu$Jy beam$^{-1}$ shows the size of the dust cavity and extent of the disk at this wavelength.  The synthesized beam of the continuum image is shown in the bottom left corner. Center: A zoom-in on the same continuum image with contours of 5, 10, 20, 25, 30, 50, and 75 times the rms noise. Key features are labeled, including the proposed co-rotational material from \citet{Long+2022} and some emission identified within the cavity and described more fully in the text. Additionally, the locations of the proposed LkCa~15~b and c \citep{Sallum+2015} are marked with a white $+$ and white $\times$, respectively. The red $+$ and red $\times$ denote instead where the planets would be 5 years later ($\sim$ the difference in time between \citet{Sallum+2015} and ALMA epochs), assuming Keplerian rotation and circular orbits. The central source is marked with a white star. Right: The same as the middle panel but using the \textit{robust}=0.9 image and its rms noise (10.4 $\mu$Jy beam$^{-1}$). Contours are 3, 5, 10, 20, 25, 30, 50, and 75 times the rms noise. The beam size of 0.055\arcsec$\times$0.031\arcsec\ is shown in the bottom corner. More faint emissions can be seen in the cavity in this image due to the higher sensitivity provided by the weighting scheme. A bridge of emission connecting the central source to the disk is also revealed.}
    \label{fig:cont}
\end{figure*}

We analyze ALMA Band 7 data of LkCa~15 taken at two different epochs. 
The first observations were obtained as part of ALMA Cycle 6 in July 2019 (Project ID:2018.1.00350.S) in an extended configuration with baselines ranging from 91~m to 8.5~km. 
The total integration time on source was about 3 hours. 
The continuum data from these observations were initially published in \citet{Long+2022}. 
LkCa~15 was observed again in October and November 2021 and May and September 2022 as part of the exoALMA program \citep{Teague_exoALMA} in a more compact configuration with baselines ranging from 10~m to 3~km.  In this case, the integration time on source was about 7.5 hours. The data were combined to increase sensitivity and reduce the effects of spatial filtering by utilizing the short baselines while retaining the high angular resolution achievable with the more extended configuration. The exoALMA-only continuum data of LkCa 15 are presented in \citet{Curone_exoALMA}. In both observations, the ALMA correlator was configured to observe both continuum emission ($\lambda = 870$ $\mu$m) and the $^{12}$CO J=3-2 transition. At the wavelength of the continuum observations, the ALMA configurations of the combined cycle 6 and exoALMA data provide a maximum resolution of about 0.02\arcsec\ and recover spatial scales as large as 18\arcsec. These correspond to spatial scales between 3.3-2800~au at the distance of LkCa~15 

The observations at each epoch were calibrated using the ALMA pipeline and then independent self-calibration was performed. 
When combining the data, there was a small shift of about 20-30 mas due to the proper motion between data acquired in 2019 and 2021 after imaging. 
To align the data, we first imaged the Cycle 6 data and created a model that was used to adjust the phase of exoALMA data to match that of the Cycle 6 data. 
After that, we performed a relative flux calibration following the procedure used in the DSHARP program as discussed in \citet{Andrews+2018}. 
For the CO observations, we applied the same calibration table generated from the continuum to calibrate and align the data. We then rescaled the flux of the CO using the scaling factor derived from the continuum. 

The exoALMA data have much higher spectral resolution than the Cycle 6 data (0.026~km s$^{-1}$ vs 0.85 km s$^{-1}$ respectively). While we lose spectral resolution when combining the data, our goal is to characterize the dust and gas emission within the inner cavity of LkCa15, which requires as high angular resolution as possible. Because of this, we chose to degrade the spectral resolution to that of the Cycle 6 data. Our final spectral resolution is 0.85 km s$^{-1}$.

We used the CASA task \texttt{tclean} to concatenate and image the continuum and molecular data applying a Briggs weighting scheme with varying \textit{robust} parameter. We find that for the continuum, a \textit{robust} parameter of 0.7 works best to achieve a good balance between signal-to-noise and spatial resolution. For the same reasons, we utilize a \textit{robust} parameter of 0.2 for the CO line. With these robust parameters, we achieve a synthesized beam size of 0.049\arcsec$\times$0.028\arcsec\ (7.8~au$\times$4.4~au) and rms noise of 11 $\mu$Jy beam$^{-1}$ for the continuum and a beam size of 0.047\arcsec$\times$0.037\arcsec\ (7.5~au$\times$5.9~au) and rms noise of 660 $\mu$Jy beam$^{-1}$ per channel for the CO.


\section{Observations} 
\label{sec:obs}

\subsection{Continuum emission} 
\label{subsec:cont}

The left panel of Figure \ref{fig:cont} shows the inner $\sim$300 au of the continuum image where a three-ring morphology previously discussed in \cite{Long+2022} is visible. 
The inner ring is the faintest of the three, while the central ring is the brightest. 
The inner ring has a peak brightness of 0.37 mJy beam$^{-1}$ and an azimuthally averaged brightness of 0.22 mJy beam$^{-1}$. 
The central ring has a peak brightness of 1.01 mJy beam$^{-1}$ and an azimuthally averaged brightness of 0.7 mJy beam$^{-1}$. 
The outer ring has a peak brightness of 0.6 mJy beam$^{-1}$ and an azimuthally averaged brightness of 0.42 mJy beam$^{-1}$. 
In addition, a central marginally resolved source with a peak brightness of 0.09 mJy beam$^{-1}$ is detected and marked with a white star in the central panel of Figure \ref{fig:cont}. 
This panel shows the inner $\sim$100 au of the disk with key features labeled. 
The most prominent of these features is the depression in the inner ring noted by \citet{Long+2022} and labeled ``Deficit". \citet{Long+2022} claim that this depression may be caused by a planet, with the higher intensity portions of the inner ring (labeled with ``Clump" and ``Arc") being co-rotational material trapped at Lagrangian points of the planet. 
The arc, clump, and deficit are all along the ring with a radius of about $\sim$43 au.

Within the 5$\times$rms continuum contour, we detect additional continuum emission. 
The morphology of this emission varies depending on the robust parameter we use for imaging (Appendix~\ref{sec:rob}). 
In the middle panel of Figure \ref{fig:cont}, we show the continuum image with robust 0.7, which reveals a compact source localized on the west side of the continuum cavity with flux density 5$\times$ higher than the rms noise (labeled ``Compact Source" in Figure \ref{fig:cont}). 
An image obtained with robust 0.9 (right panel of Figure~\ref{fig:cont}) reveals that this compact source is surrounded by diffused emission (labeled ``Faint Emission") extending along the nearest side of the disk (i.e., the northwest side). 
This image also reveals a bridge of emission connecting the central source to the innermost ring along the southeast direction. 
\citet{Facchini+2020} also detected faint continuum emission within the main rings but, due to the lower sensitivity of their observations, they concluded that it could just have been an imaging artifact. 
The Faint Emission appears to be spatially coincident with the bulk of the scattered light images \citep{Thalmann+2016, Ren+2023}. However, none of the mm-wave emission peaks correspond to the location of the candidate planets LkCa~15~b and c ($+$ and $\times$ in middle panel of Figure \ref{fig:cont}). 
These findings suggest that the Faint Emission might trace an even fainter and perhaps asymmetric dust ring characterized by a radius of about 15-20 au.
This is in agreement with the results of \citet{Sallum+2023} which claim that the scattered light and H$\alpha$ in the inner disk can be explained by a dynamic small grain population.

Finally, the central emission appears mostly point-like with a shape of the 5$\sigma$ contour of the same size and position angle as the synthesized beam. 
However, at lower contour levels and in the images obtained with higher robust values, the central emission extends toward the southeast, and it connects to the inner continuum ring.  

\begin{table}[t]
    \centering
    \begin{tabular}{lllll}
    \hline
    \hline
    Feature & r$_{x}$ & r$_{x}$ & $w_x$ & $I_x$ \\ 
            &(mas)    & (au)  &   (au)    & (mJy/$\arcsec^{2}$) \\
    \hline
    CS &    0            &    0           & 1.47$^{+0.7}_{-0.6}$  & 287$^{+486}_{-143}$ \\
    B43  & 272.4$^{+1.7}_{-1.6}$  & 43.3$\pm$0.3   & 8.4$\pm0.4$            & 115$\pm3$ \\
    B69  & 432.7$\pm0.4$          & 68.8$\pm 0.1$  & 7.67$\pm0.1$           & 478$\pm3$\\
    B100 & 629.8$^{+0.8}_{-0.9}$  & 100.1$\pm 0.1$ & 14.1$\pm0.2$           & 235$\pm2$ \\
    \hline
    \end{tabular}
    \caption{\label{tab:table} Properties of the dust emission components modeled as Gaussian functions with intrinsic radial profiles $I(r) = I_x e^{-(r-r_x)^2/2w_x^2}$ (Appendix~\ref{sec:ring_geo}). The central source (CS) and the dust rings (B) are centered at RA=04:39m:17.81s+(-0.0404\arcsec$\pm$0.0003\arcsec) and Dec=+22:21:03.13`+(-0.0844\arcsec$\pm$0.0002\arcsec). The disk inclination and position angle are $50.7\arcdeg \pm 0.1\arcdeg$ and $61.7\arcdeg \pm 0.1\arcdeg$, respectively.}
\end{table}

Table \ref{tab:table} lists the radii, widths, and intensity of the three dust rings, as well as the width and intensity of the central source, obtained by modeling the ALMA observations in the {\it uv} plane using {\it Galario} \citep[][see Appendix~\ref{sec:ring_geo}]{Tazzari+2018}.   
The dust rings share the same center, inclination (50.7\arcdeg$\pm$0.1\arcdeg), and position angle (61.7\arcdeg$\pm$0.1\arcdeg), implying that they are intrinsically circular, concentric, and co-planar. 
Furthermore, they are almost equally spaced in radius, with $r_3/r_2 \simeq 1.5$ and $r_2/r_1 \simeq 1.6$.
Interestingly, these ratios are very close to a 2:1 mean motion resonance. 
All three rings appear to be spatially resolved in the radial direction with relative widths $w_x/r_x$ of 0.19, 0.11, and 0.14 for B43, B69, and B100, respectively. 
The implications of these ratios in terms of dust trapping are discussed in Appendix~\ref{sec:ring_geo}.

Finally, we show in Figure~\ref{fig:Galario} an image obtained by subtracting the galario best fit model, whose details are outlined in Table~\ref{tab:table}, from the observations (Appendix~\ref{sec:ring_geo}). 
The Clump and the Arc are visible as positive residuals. The Deficit appears instead as a negative residual.
Additionally,  a large positive arc-like residual occurs on the western side of the B69, while the eastern side is dimmer relative to the model. 
These asymmetries in the continuum emission, which likely trace azimuthal variations in the dust density, are important for interpreting the observations regarding planet-disk interaction models and will be further discussed in the next section.

\begin{figure} [!t]
    \centering
    \includegraphics[width=\linewidth]{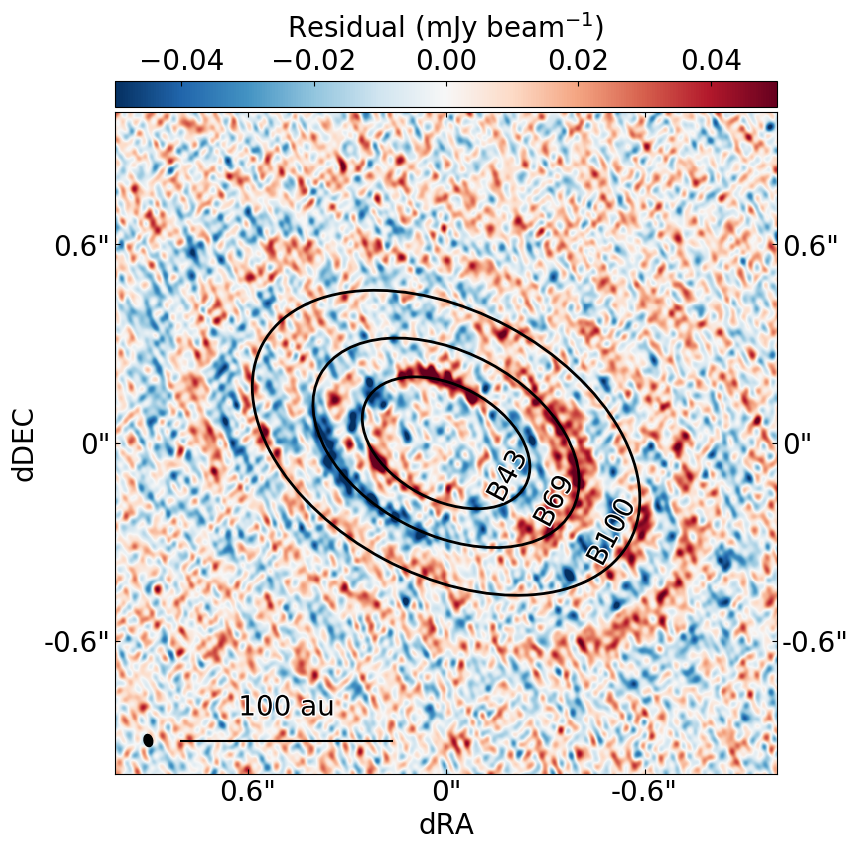}
    \caption{
    Residual map obtained by subtracting Galario's best-fit model from the observations in the Fourier space. The central emission is removed  by the model. The locations of the bright rings found in galario are marked with ellipses. Features identified in Figure \ref{fig:cont} are visible here. Both Clump and Arc are visible along the inner ring (B43) as dark red residuals while the deficit has some of the most negative values. The bright continuum ring (B69) also shows asymmetries, being brighter in the west and fainter in the east.}
    \label{fig:Galario}
\end{figure}

\subsection{CO emission} \label{subsec:CO}

\begin{figure*}
    \centering
    \includegraphics[width=\linewidth]{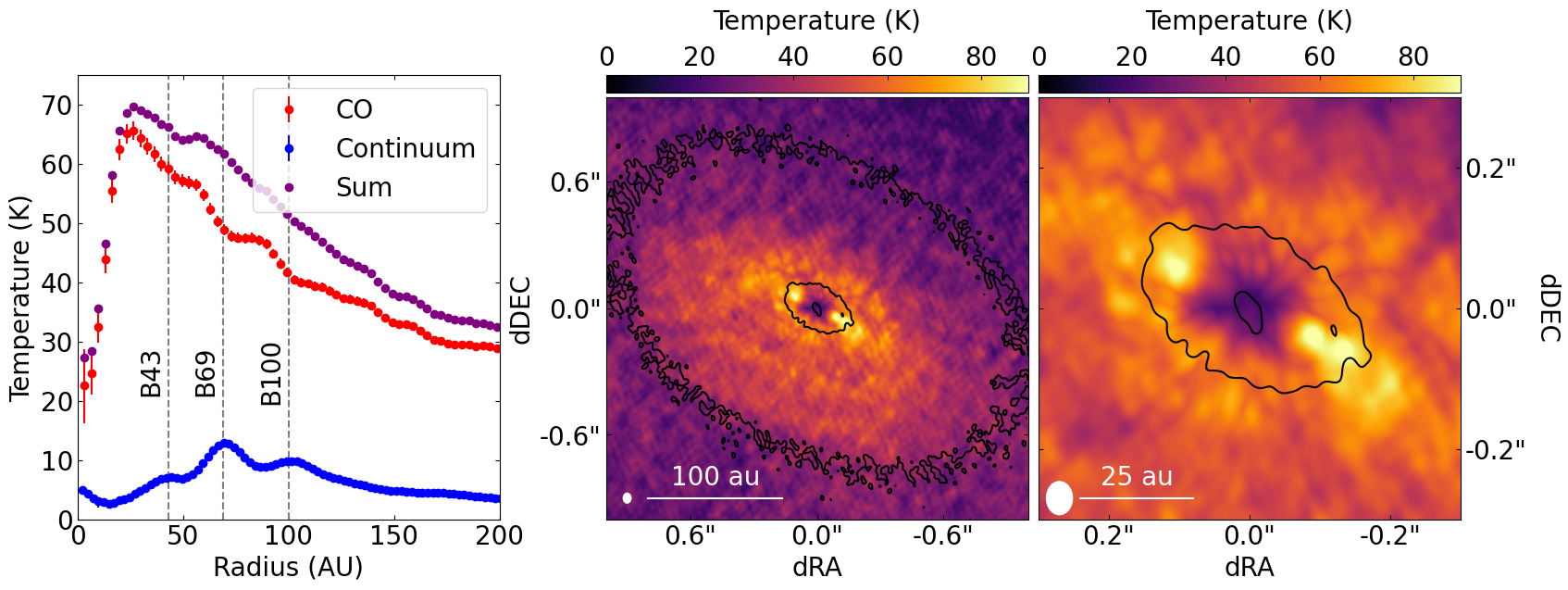}
    \caption{Left: azimuthally averaged radial brightness temperature profile of both the continuum-subtracted CO and continuum. The sum of the two profiles in this panel is also shown to highlight the effects of the subtraction. The peaks in the continuum profile correspond to drops in the CO, showing that the continuum is likely optically thick at the location of the prominent rings. The rings are marked with vertical dashed lines. Center: moment 8 peak intensity map of the CO converted to brightness temperature for the inner 2\arcsec$\times$2\arcsec. The contour value is again 5 times the continuum rms and shows the continuum cavity and central source. The synthesized beam size is shown in the bottom left corner. Right: same as the central panel but zoomed into the inner 0.6\arcsec$\times$0.6\arcsec to show the extent of the CO emission within the cavity.}
    \label{fig:moment}
\end{figure*}

Figure \ref{fig:moment} presents the continuum-subtracted $^{12}$CO J=3-2 peak brightness temperature (Using the full Planck law)\footnote{To improve the quality of the peak intensity map, we generated and coadded four CO data cubes, each offset in velocity by 0.17 km/s. This procedure largely removes image artifacts generated by the coarse velocity resolution when taking the peak intensity of each pixel and does not affect the physical interpretation of the data.}. In terms of brightness temperature, the rms noise of the CO data is 6.8~K while for the continuum it is 2.6~K.\footnote{Note that the brightness temperature using the full Planck law does not scale linearly as does the flux density. The values listed here correspond to the 1$\sigma$ rms values of the flux density.} A figure with the CO channel maps is available in Appendix~\ref{sec:co_channel}. 
Overall, the CO emission covers a wide range of spatial scales, extending both outside the continuum emission and inside the dust cavity. 
As a reference, we overlay the 5$\sigma$ contour of the continuum emission. Additionally, the CO emission is more asymmetric than the continuum along the azimuthal direction. 
In particular, the CO emission in the southeast half of the disk is brighter and extends farther away from the center than the opposite side. 
This is due to the flaring of the CO-emitting layer, which causes the nearest side of the disk (the northeast side in this case) to be compressed in the direction perpendicular to the line of sight. 

In the left panel of Figure~\ref{fig:moment} we plot both continuum and CO profiles, as well as their sum. Dashed lines indicate the locations of the bright continuum rings whose subtraction leaves clear signatures in the CO profile. This shows that the continuum is likely optically thick at these radial locations. Interestingly, there are modulations seen in the outer disk of both the CO profile and the sum profile. These may also be associated with continuum features \citep[see e.g. Fig. 2 of][]{Curone_exoALMA}, or may be driven by other mechanisms \citep{Longarini_exoALMA,Barraza_exoALMA}.

While the continuum emission peaks at 68 au, the CO emission peaks at 25 au, well within the innermost dust ring and the orbital radius of the candidate exoplanets LkCa~15 b and c.
Taken at face value, the lack of any cavity in $^{12}$CO seems to disfavor the presence of massive planets in the disk's innermost regions, which would instead clear out both dust and gas and trap dust at pressure maxima \citep{Zhu+2012}. 
However, the different radial profiles might be caused by radiative transfer effects, such as the different vertical distribution and optical depths of dust and CO emission. 
For these reasons, it is important to analyze the observations using theoretical models that include radiative transfer calculations of the dust continuum and CO emission. 


\section{Hydrodynamic and Radiative Transfer Models} \label{sec:mod}

Our main goal is to investigate if the different morphology of continuum and CO emission observed within LkCa~15's disk dust cavity might be consistent with the presence of massive planets and derive constraints on their mass if this is the case. 
We do this by comparing ALMA observations with hydrodynamic+radiative transfer simulations of the planet-disk interaction for models characterized by different planetary configurations. 
However, a full exploration of the model parameter space is impossible due to the large computational time required to perform such simulations.
Instead, we start with simple assumptions on the number and radial distribution of planets, and the gas and dust properties, and derive robust conclusions against the specific choice of parameters.  


The general modeling procedure is as follows: using some initial profiles (e.g. surface density, aspect ratio, etc., see Section~\ref{subsec:setup}), we define an initial disk. With the disk defined, we can then use radiative transfer codes to calculate the temperature of the initial disk and hydrodynamic codes to evolve the density perturbations created by embedded planets. Note that we calculate the disk temperature only for the initial unperturbed disk state and assume it to remain constant with time. 
This might not be a good assumption if planets carve deep gaps in the disk \citep{Facchini+2018b}, but is the only option for exploring a wide range of model parameters in the absence of a fast radiation-hydrodynamic code\footnote{Running the temperature calculation on a perturbed disk with deep gaps results in increases on the order of only a few percent throughout most of the disk. Larger changes occur in the vertical transition region and the gaps, and reach a maximum of $\sim50\%$}. As an additional caveat we calculate only the dust temperature, assuming that it is equal to the gas temperature. The implications of this assumption will be discussed further in Section~\ref{subsec:model:CO} Armed with a temperature, dust and gas densities, and gas velocities, we can then produce synthetic dust continuum and molecular line images, taking into account the finite angular and spectral resolution of the observations. The details of the codes used and the initial disk set up are described below.

\subsection{Model Setup} \label{subsec:setup}

We use the 2D ($r, \phi$) version of the hydrodynamic code LA-COMPASS \citep{Li+2005,Li+2009} to calculate the evolution of circumstellar gas and dust under the gravitational influence of planets. 
Following the evolution of an $\alpha$ disk \citep{Lynden-Bell+1974,Pringle+1981,Hartmann+1998,Isella+2009}, we assume an initial gas surface density profile
    \begin{multline}
        \Sigma_g(r) = \Sigma_{c}\left(\frac{r}{r_c}\right)^{-\gamma} \\
        \times\exp\left\{-\frac{1}{2(2-\gamma)}\left[\left(\frac{r}{r_{c}}\right)^{(2-\gamma)}-1\right]\right\},
    \end{multline}
where $\Sigma_{c} = 1.62$ g cm$^{-2}$ is the surface density at the cutoff radius $r_{c} = 150$ au and $\gamma$ is the slope of the disk surface density which we set to 1. In our models, we assume a radially constant value of the viscosity parameter $\alpha = 2 \times 10^{-3}$.
The initial values of $\Sigma_c$ and $r_c$ are based on the results of \cite{Facchini+2020} and correspond to an initial disk mass of 0.04 M$_\odot$. We did not include disk self-gravity in our models as these profiles imply stability under the Toomre Q parameter \citep{Toomre+1964}. Initially, dust and gas are perfectly mixed with a gas-to-dust ratio of 100. Whereas LA-COMPASS allows for the inclusion of multiple dust sizes, we assume that all the dust is in grains with a size of $a = 0.14$~mm and density $\rho_s = 1.26$ g cm$^{-3}$ \citep{Facchini+2020}. At the cutoff radius of 150 au, this corresponds to St = 0.017.
This value of $a$ was chosen as the dust grain size which carries the maximum opacity at a given wavelength goes as a $\sim\frac{\lambda_{\text{obs}}}{2\pi}$. Therefore, 0.14~mm grains should dominate the continuum emission observed at 0.87~mm.  

We perform hydrodynamic simulations assuming a locally isothermal equation of state. 
The temperature profile of the disk follows the power law relation 
\begin{equation}
     T(r) = 41.6\text{ K}\left(\frac{r}{25\text{ au}}\right)^{-1/2},
\end{equation}
which resulted from the radiative transfer calculation discussed below.      
In a vertically isothermal disk in hydrostatic equilibrium, the assumed temperature profile can be equivalently expressed in terms of the disk aspect ratio as
    \begin{equation}
        \frac{H}{r} = 0.058\left(\frac{r}{25 \text{ au}}\right)^{0.25}, 
    \end{equation}
where $H$ is the disk pressure scale height given by $\frac{H}{r}$=$\frac{c_{s}}{v_{k}}$. In this case, $v_{k}$ is the Keplerian velocity, and $c_{s}$ is the sound speed. Our simulation domain covers from 4 to 200 au in the radial direction and 0 to $2\pi$ in azimuth. While other exoALMA papers \citep{Galloway_exoALMA} show that there is $^{12}$CO emission out to $\sim600$ au, we are only concerned with the inner disk structures, thus motivating our choice of outer boundary. We use 1024 logarithmically spaced cells in the radial direction and 3072 uniformly spaced cells in azimuth. We tested that both the chosen boundary conditions and grid resolution do not affect the results of the simulations. Ultimately, we use open boundaries as this led to the best agreement with the steady state, similarity solution for the evolution of a viscous disk over time \citep{Isella+2009}. 

\begin{figure*}[!t]
    \centering
    \includegraphics[width=0.9\linewidth]{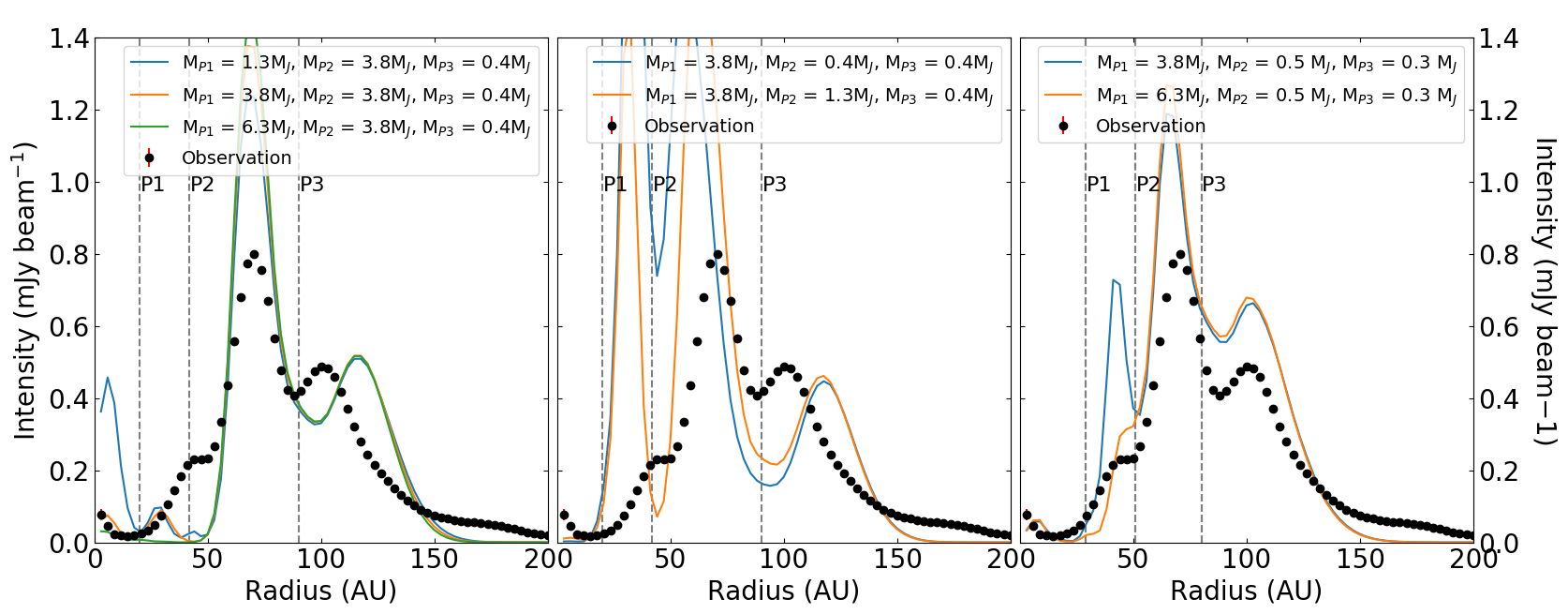}
    \caption{Comparing the azimuthally averaged radial intensity profiles from our models to the observations. In all panels the filled circles show the radial profile of the dust continuum emission while colored lines show profiles from the models. Vertical dashed lines show the orbital radii of the three planets in the model. Left: continuum models with the lowest $\chi^{2}$ values from our model grid characterized by fixed planets' orbital radii. Middle: reducing the mass of Planet 2 results in too much dust emission within 70 au. Right: relaxing the assumption of fixed radii allows for the planets to be placed at locations that better reproduce the observations.}
    \label{fig:mod_cont_rad}
\end{figure*}


To calculate the disk temperature and the emerging emission in the dust continuum and CO line emission, we use the publicly available 3D radiative transport code RADMC3D \citep{Dullemond+2012}. 
Under the assumption of a passively irradiated disk, the disk temperature depends on the incoming stellar radiation, on the dust opacity (we assume that atomic and molecular gas play negligible roles in setting the disk's interior thermal structure), and on the dust vertical and radial distribution. 
For the irradiation flux, we assume a black body spectrum at a temperature $T_\star=$ 4500~K and total luminosity $L_\star=$ 0.95~L$_\odot$ \citep{Donati+2019}. For the dust opacities, we assume the DSHARP total absorption opacities discussed in \citet{Birnstiel+2018} (see equation 6), corresponding to a size-weighted opacity where the grain size which dominates the absorption at a given wavelength is $a\sim\frac{\lambda_{\text{obs}}}{2\pi}$. This gives us an absorption opacity at each wavelength which no longer depends on the dust size, thus allowing for the approximation of one dust size in our hydro models. 
The vertical gas density profile is calculated assuming that the disk is vertically isothermal and in hydrostatic equilibrium, which leads to
\begin{equation}
    \rho_g(r,\phi,z) = \frac{\Sigma_g(r,\phi)}{H\sqrt{2\pi}}\exp\left[-\frac{z^{2}}{2H^{2}}\right],
\end{equation}
where $\Sigma_g$ is the gas surface density calculated by LA-COMPASS. 
The vertical dust density profile is calculated as
\begin{equation}
    \rho_d(r,\phi,z) = \frac{\Sigma_d(r,\phi)}{H_d\sqrt{2\pi}}\exp\left[-\frac{z^{2}}{2H_d^{2}}\right],
\end{equation}
where $\Sigma_d$ is the dust surface density calculated by LA-COMPASS and $H_d = H_g \times \left( 1 + \frac{\textrm{St}}{\alpha} \right) ^{-1/2}$ is the pressure scale height of the dusty disk which accounts for dust settling toward the midplane \citep{Youdin+2007}. 
In the last equation, St is the dust Stokes number calculated as $\textrm{St} = \frac{\pi \rho_s a}{2 \Sigma_g}$. Including dust settling is important to model ALMA observations and avoid the mm-wave emitting layer artificially extending in the vertical direction, particularly when observing inclined disks. 

The disk temperature results from a balance between heating (provided by the direct stellar irradiation via the dust opacity and reprocessed radiation from dust itself) and cooling (controlled by dust opacity). 
These processes primarily occur at optical and infrared wavelengths and are controlled by the distribution of (sub)micron-size grains that carry most of the opacity at these wavelengths. 
In first approximation, these grains have $\textrm{St}/\alpha \ll 1$, meaning that they are well coupled to the gas. For this reason, to calculate the disk temperature, we assume $\rho_d = 0.01 \rho_g$ and a dust opacity corresponding to a grain size distribution with a maximum grain size of $1 \mu$m. 
To find the disk temperature profile corresponding to the hydrostatic equilibrium, we start from an initial guess of the disk aspect ratio and iterate on the temperature calculated by RADMC3D until convergence. 
Using the gas surface density profile of Equation 1, this procedure leads to the midplane temperature profile reported in Equation 2 and the disk aspect ratio of Equation 3.

After the temperature calculation, we use the ray tracing capabilities of RADMC3D to generate synthetic continuum and CO emission images. 
For the continuum images, we use the surface density of the 0.14~mm dust grains calculated by LA-COMPASS together with the corresponding DSHARP opacity and the disk inclination and position angle inferred in the previous section (see Table 1). 
For the CO images, we utilize the gas density and velocity calculated by LA-COMPASS. 
To generate the 3D CO density distribution, we assume a CO/H abundance ratio of $5\times 10^{-5}$ and incorporate both CO freezeout and photodissociation following the prescriptions outlined in \citet{Rosenfeld+2013}, \citet{Qi+2011}, and \citet{Aikawa+1999}. 
Calculating the photodissociation surface involves integrating the vertical gas distribution from the top down and finding the value of $z = z_{phot}$ at which it reaches a threshold value $\sigma_{phot} = 5\times10^{20}$ cm$^{-2}$ \citep{Visser+2009}. This is by no means a complete treatment of photodissociation in protoplanetary disks, and we refer the reader to Section \ref{subsec:model:CO} and Appendix~\ref{sec:photodis} for a more complete discussion on the implementation. 
For CO freezeout, we set the CO abundance to zero where the gas temperature is below 19 K \citep{Rosenfeld+2013}.
We ray-trace the CO emission by oversampling the velocity resolution of the observations. We calculate the CO emission across 289 channels at a resolution of 0.05 km s$^{-1}$, and then average it down to the observed resolution of 0.85 km s$^{-1}$. The CO simulations include dust opacity to account for continuum absorption properly. Finally, to compare to the continuum-subtracted CO observations, we calculated a continuum-only data cube and subtracted it from the CO synthetic channel map.   

\subsection{Continuum Modeling}
\label{subsec:model:cont}

The models of the planet-disk interaction depend on many parameters, and exploring all of them in a systematic way is unfeasible. 
Instead, we adopt a step-by-step approach based on our observations and previous results and use our intuition to optimize the model parameters to reproduce the observations.   


All of our models include three planets. Our initial assumption is that the first planet (Planet 1) has an orbital radius of 20 au, corresponding to the stellocentric distance of the candidate planets LkCa~15 b and c, as well as, the compact source discussed in Section~\ref{sec:obs}.  
Our working hypothesis is that Planet 1 is responsible for clearing out the innermost disk regions. As the existence of multiple planets in the cavity remains in doubt, we attempt to create a cavity with only one massive planet to reduce our parameter space. 
The second planet (Planet 2) is located at an orbital radius of 42 au, corresponding to the radius of the dust deficit and inner ring. 
This follows the hypothesis of \cite{Long+2022} that the dust emission observed along the innermost ring arises from dust corotating with a planet. 
The third planet (Planet 3) has an orbital radius of 90 au, corresponding to the location of the outermost gap. 
In this case, we assume that the planet is responsible for clearing the gap seen in the continuum emission.
Keeping the planets' orbital radii fixed, we then run a grid of models allowing for three values for the mass of each planet. 

For all of our models we assume a stellar mass of 1.25 M$_{\odot}$ \citep{Donati+2019}. Due to the large cavity in the continuum, one would expect Planet 1 to be massive. Planets 2 and 3 are likely less massive given the dust morphology of the outer disk. An estimate of the gap-opening criteria from \citet{Dipierro+2017} indicates that planet masses of 3.8, 1.9, and 0.8 M$_{\textrm{J}}$ will open gaps in the dust at the chosen radii of the planets (20, 42, and 90 au respectively). 
Because of this, we chose 1.3, 3.8, and 6.3 M$_{\textrm{J}}$ as possible values for planet 1 mass. 
For planets 2 and 3, we chose masses of 0.4, 1.3, and 3.8 M$_{\textrm{J}}$. These choices for the planet masses allow us to run models both above and below the theoretical gap-opening criteria.
We run each simulation for 0.625 Myr (5000 orbits at 25 au). This allows for the simulations to achieve a quasi-steady state, meaning that the results do not strongly depend on the time frame used for the comparison. For all of our simulations, we assume that these planets are at fixed radial locations on circular orbits and we do not vary the azimuthal angle at which they are initially placed.

For a quantitative comparison between models and observations, we use the {\it chi2image} function from {\it galario}, which calculates a $\chi^{2}$ by comparing synthetic images and the observed continuum emission in the visibility space. In this way, we naturally account for the angular resolution and discrete {\it uv} sampling of ALMA observations. 

The left and middle panels of Figure~\ref{fig:mod_cont_rad} summarize the main results. Overall, none of these models provide a good match to the observations. 
The models with the lowest $\chi^2$ (left panel) have $M_{p1}>1.3$~ M$_\textrm{J}$, $M_{p2}=3.8$~ M$_\textrm{J}$, $M_{p3}=0.4$ ~M$_\textrm{J}$. 
Planets 1 and 2 clear out material in the cavity and produce a bright continuum ring at the location of the main dust ring seen in the observations. 
However, due to the high mass, Planet 2 creates a large gap at the location of the innermost ring in the continuum. Lowering the mass of Planet 2 (middle panel of the figure) leads to more continuum emission at the location of the inner ring but results in the appearance of an undesired ring of dust between Planet 1 and 2 at a radius of about 35 au. 
An additional feature of these models is that the dust ring created by Planet 3 has a larger radius than the outermost dust ring. 

\begin{figure*}[!t]
    \centering
    \includegraphics[width=\linewidth]{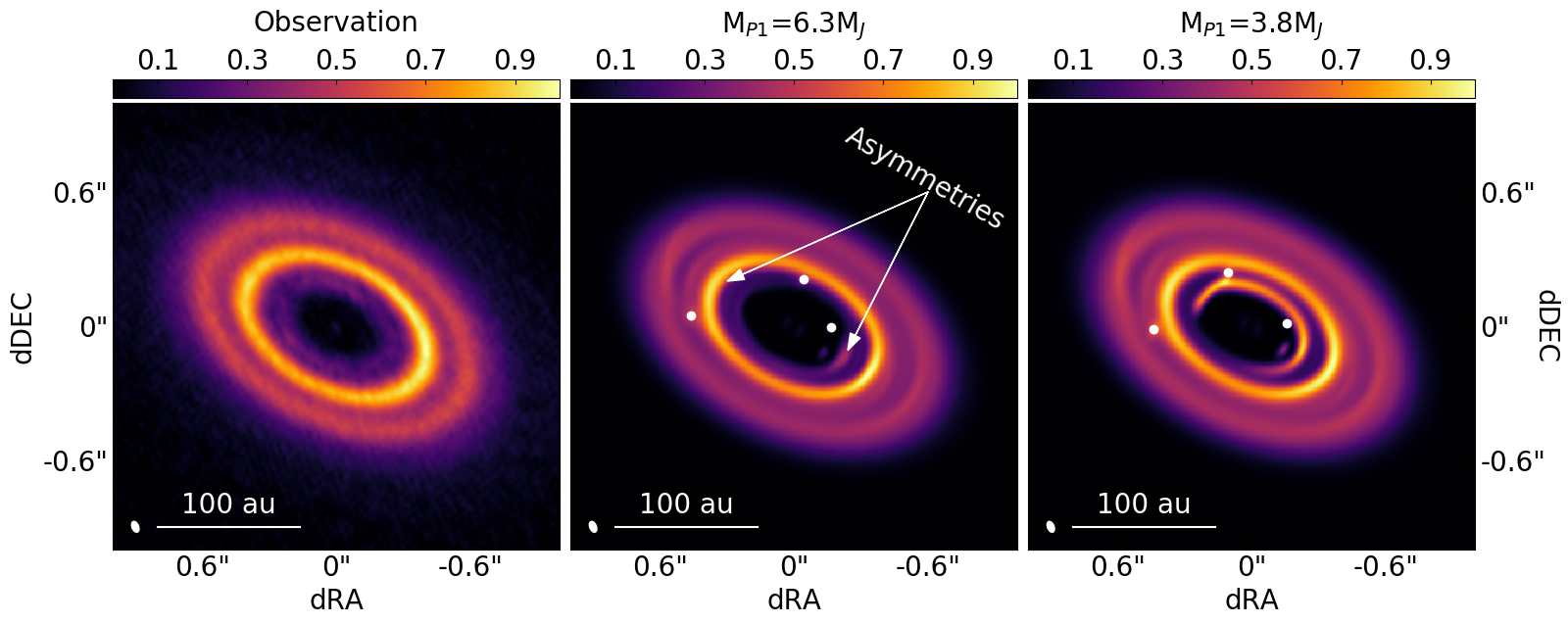}
    \caption{Left: continuum map of LkCa~15 as shown in Figure \ref{fig:cont}. Center and Right: synthetic continuum images corresponding to the  models that provide the best match to the observations (see also right panel of Figure \ref{fig:mod_cont_rad}). In each panel, the continuum intensity has been normalized to the peak intensity to facilitate the comparison between models and observations in terms of the level of azimuthal asymmetries. We highlight a few of the asymmetries seen in the inner and middle rings. The locations of the planets are shown with white dots in the models.}
    \label{fig:mod_cont_im}
\end{figure*}

Following these results, we relaxed our assumptions on the planets' orbital radii and tried several combinations of planets' masses and orbital radii to reproduce the observed continuum radial profile. 
We increased the orbital radii of Planet 1 and 2 to match the morphology of the dust cavity and moved Planet 3 inward to match the radius of the outermost ring. 
We find the best agreement in the locations of the gaps and rings between models and observations when Planet 1 has an orbital radius of about 29 au and a mass between 3-5 M$_\textrm{J}$, Planet 2 has an orbital radius of 51 au and a mass of 0.4 M$_\textrm{J}$, and Planet 3 has an orbital radius of 80 au and a mass of 0.25 M$_\textrm{J}$ (see right panel of Figure~\ref{fig:mod_cont_rad} and Figure~\ref{fig:sigma_gas}). 
In these models, Planet 1 is responsible for the large dust continuum cavity, while the innermost dust ring consists of dust trapped between Planet 1 and 2.
The models reproduce the location of the dust rings and the overall morphology of the dust cavity, but over predict the intensity profile across the middle and outer rings by 30\%-50\%. 
Additionally, a massive Planet 1 generates a steep increase in the dust intensity profile between 30-40 au, while the observations show a more gradual increase. 
Matching the exact radial profile of the dust continuum emission is beyond the scope of our investigation.
However, we note that lower continuum intensities could be obtained, for example, by adopting a different (lower) mm-wave dust opacity. 
Additionally, while the true disk temperature may differ slightly from the adopted model which was calculated for the unperturbed disk \citep{Isella+2018}, the midplane values are largely in agreement with those derived in \citet{Galloway_exoALMA}. 

Figure \ref{fig:mod_cont_im} shows synthetic observations of the two models presented in the right panel of Figure~\ref{fig:mod_cont_rad}. To better compare the azimuthal structure of the synthetic emission to the observations, we show the normalized intensity in the figure, keeping in mind the differences in absolute intensity discussed above.
Overall, the models qualitatively match the observations in that the innermost and middle rings are asymmetric, while the outermost ring is mostly symmetric. Notably, the middle rings are characterized by large-scale azimuthal bumps that match the asymmetry discussed in Section~\ref{sec:data}. 
All our models also show compact emission at the location of the planets.
However, this feature is caused by an incorrect treatment of the circumplanetary material wherein the material is expanded vertically using the scale height of the PPD rather than the CPD. Therefore, these sources of compact emission should be disregarded in the model images.
Considering all the caveats discussed above, these two models for the continuum emission match the size of the dust cavity and the level of dust depletion within the cavity. 
In particular, the model with $M_{p1} = 3.8$~M$_\textrm{J}$ reproduces the dust continuum emission within 30 au, providing a base for constraining the gas depletion in the innermost regions of LkCa~15's disk.  
In the next section, we will adopt this model as a reference model to investigate the CO emission arising from the dust-depleted cavity

\subsection{CO Modeling}
\label{subsec:model:CO}

\begin{figure*}
    \centering
    \includegraphics[width=\linewidth]{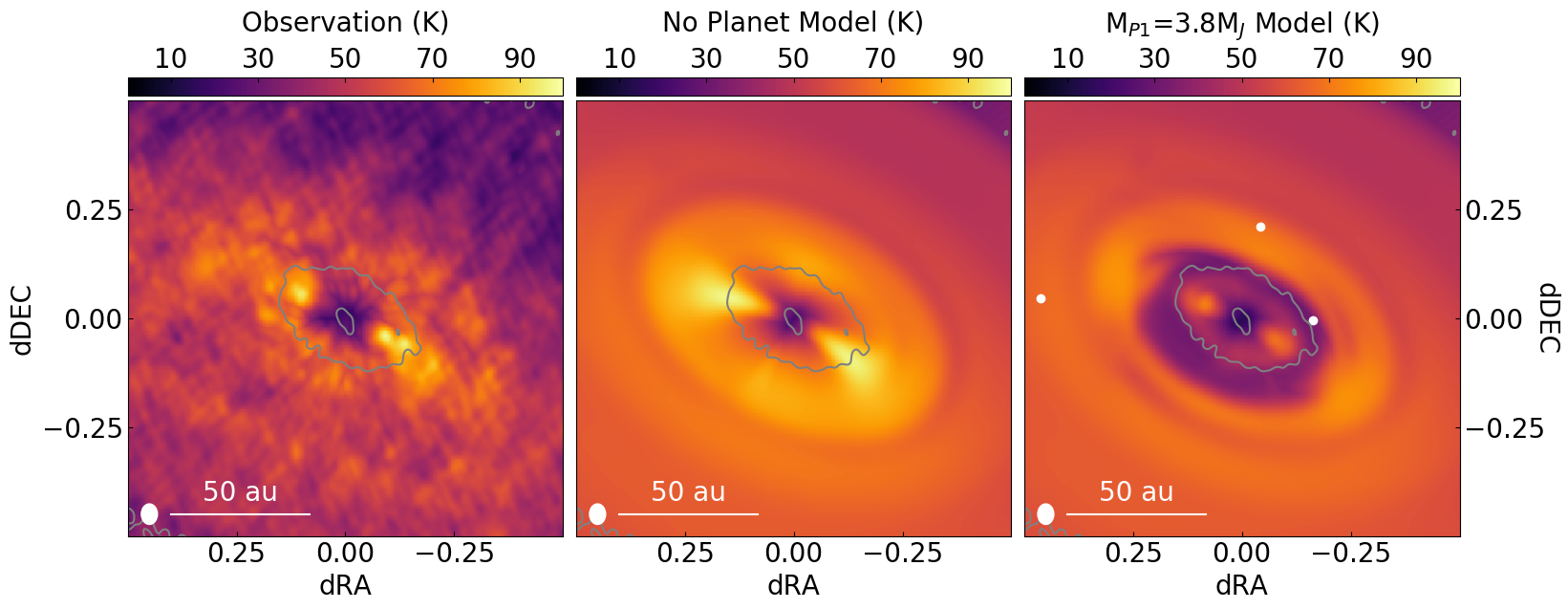}
    \caption{Left: The peak intensity map of CO emission recorded toward LkCa15 converted to brightness temperature. This is the same image shown in Figure \ref{fig:moment}. Center: CO peak intensity map of the disk with no planets included. The resemblance between this image and the observations indicates that a disk with no planets can approximate the observed CO emission. Right: CO peak intensity map of the reference model discussed in Section~\ref{subsec:model:cont} with a 3.8 M$_\mathrm{J}$ inner planet. We mark the locations of the planets in this model with white dots as in Figure~\ref{fig:mod_cont_im}. In this model, the effect of the massive planet is clearly visible in the CO emission and shows that a massive planet is not consistent with the observations. The corresponding surface density maps are shown in Figure~\ref{fig:sigma_gas}.} We overplot the 5$\sigma$ contour of the continuum in all images to emphasize the location of the dust cavity. CO photodissociation is included in both of the models shown.
    \label{fig:mod_temp_im}
\end{figure*}

Figure \ref{fig:mod_temp_im} shows synthetic observations of the CO peak intensity emission corresponding to our initial disk model with no planets, and the reference model discussed above which is characterized by three planets with masses of 3.8, 0.5, and 0.3 M$_\textrm{J}$ orbiting at 29, 51, and 80 au respectively.
CO photodissociation is included in both the middle and right panels. Adopting the values mentioned in Section~\ref{subsec:setup} for the photodissociation vertical threshold and CO abundance leads to little change in the depletion of CO within the dust cavity (see Appendix \ref{sec:photodis} and Figure~\ref{fig:mod_temp_photodissociation}).

The effects of the massive innermost planet are clearly visible in the inner disk in the model shown in the right panel of Figure~\ref{fig:mod_temp_im}, where gas has been cleared out and CO emission reduced. 
The CO emission within $\sim$0.2\arcsec\ from the center becomes optically thin in this model. 
This result suggests that if the inner disk of LkCa~15 is shaped by a single planet, its mass is likely lower than $\sim$3~M$_\textrm{J}$. Even in the case of our $\sim$1 M$_\textrm{J}$ model (with or without photodissociation), a cavity is still present in the synthetic CO images, suggesting that if there is a planet its mass must be $\lesssim 1 $ $M_\textrm{J}$. In fact, a model that does not include any planets (see middle panel of Figure \ref{fig:mod_temp_im}) matches the observed CO emission. 

To determine whether a higher gas temperature can account for the observed emission in the presence of gas depletion by a planet, we recalculate the disk temperature after the innermost planet has cleared a cavity. As a result, the gas temperature and CO emission within the cavity increase by approximately 10\%. However, this enhancement is insufficient to explain the observed emission in the presence of a massive planet.
A discussion of the implications of these results is presented in the next section. 

\begin{figure*}
    \centering
    \includegraphics[width=\linewidth]{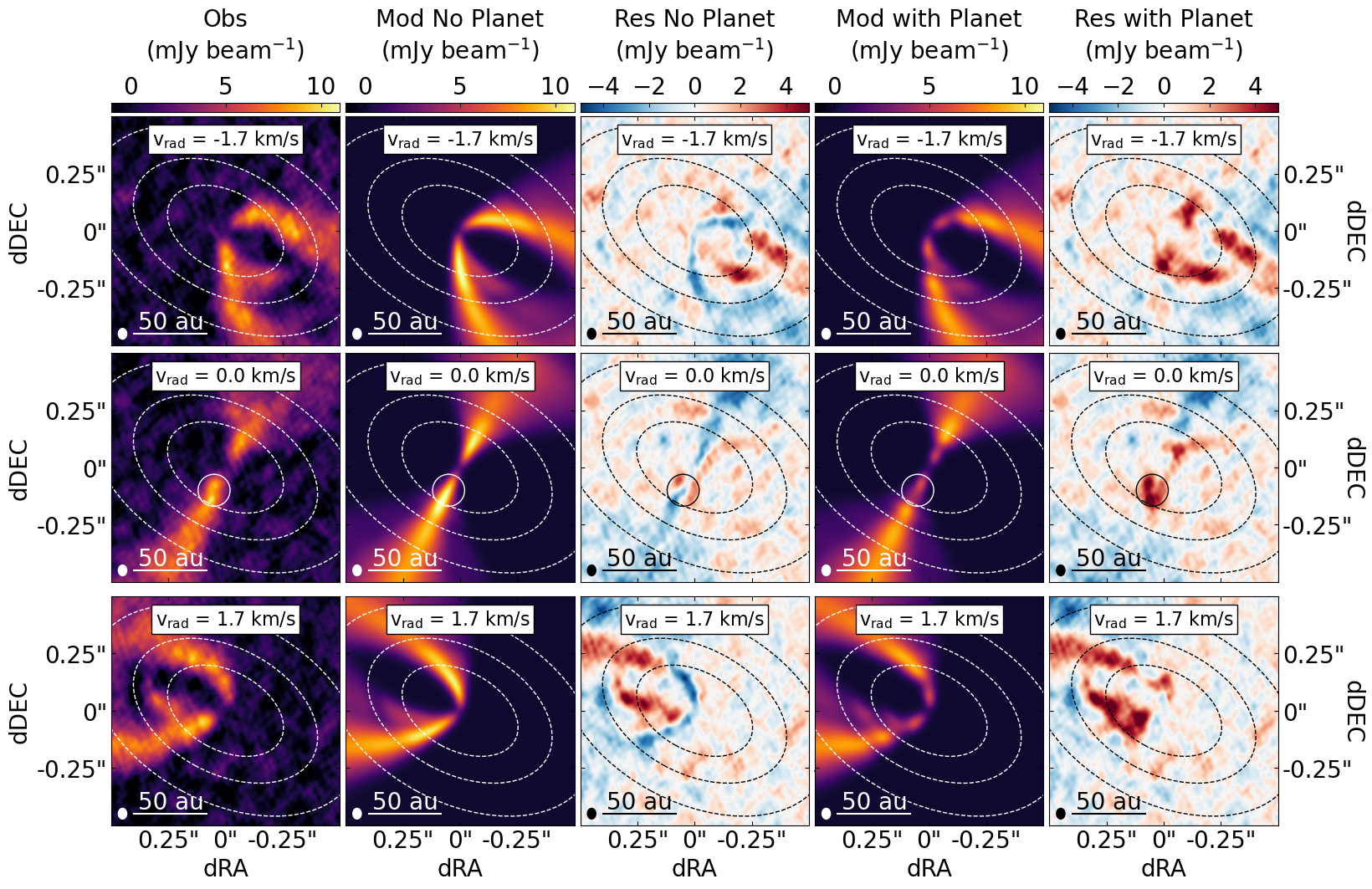}
    \caption{Select channels from the $^{12}$CO (J=3-2) data cube of the observations (1st Column), the models shown in Figure~\ref{fig:mod_temp_im} (2nd and 4th Columns), and the residuals of the observations minus the models (3rd and 5th Columns). The dashed ellipses in every panel show the location of the 3 continuum rings derived from our continuum observation analysis. Some key features of the observations are reproduced well in the models, namely the backside of the disk being visible. Additionally, both sides of the disk dim in roughly the same place as in the observations due to continuum subtraction from the bright continuum rings (see e.g. the dark lanes in CO  coincident with the location of B69). In the model shown with planets, red residuals tend to follow the B43 ring since this model over predicts its brightness and thus subtracts too much continuum. The residuals in both models show that the observations are brighter in the inner disk, while the models are brighter in the outer disk. The zero velocity channel shows a small s-shaped kink which is marked by a circle in the middle row and shows up as dark red spots in the residual maps. CO photodissociation is included in both of the models shown.}
    \label{fig:res_channel_map}
\end{figure*}

Finally, it is worth noting similarities and differences between CO synthetic models and observations that highlight the successes and shortcomings of our modeling procedure. 
These features are better seen by comparing observed and synthetic channel maps as shown in  Figure~\ref{fig:res_channel_map}.
Most importantly, our models of disks both excluding (Column 2 of Figure~\ref{fig:res_channel_map}) and including (Column 4 of Figure~\ref{fig:res_channel_map}) planets match the overall morphology of the CO emission, which is primarily controlled by the disk differential rotation, the vertical CO distribution, and CO temperature. 
For example, the CO emission observed at $\pm$1.7 km s$^{-1}$ relative to the central star velocity, shows spatially separated emission from both the front and back sides of the disk \citep{Rosenfeld+2013}.  
Importantly, the CO disk dims in a dark lane in roughly the same location in both the observations and models which corresponds to the location of the bright continuum ring and is a result of the continuum subtraction.
This is made particularly clear in the residual channels of Column 5 where the red residuals follow the B43 ring since this specific model we show over predicts the continuum brightness of that ring (see blue profile in right panel of Figure~\ref{fig:mod_cont_rad}).
Indeed, the optical depth at 0.87mm, calculated from the DSHARP absorption opacity and dust surface density of our model, shows that the disk is optically thin in dust but becomes marginally optically thick in the bright rings. 
Furthermore, the shape of the CO emission arising from the near side of the disk, which in the case of LkCa~15 is the north-west side, differs in shape from the emission arising from the opposite side, appearing more compressed and less extended than emission on the far side of the disk (see Figure \ref{fig:res_channel_map}, 0 velocity channels).
Both of the models shown reproduce these features, giving us confidence in our assumption of the stellar mass and our calculation of the disk geometry, temperature, and vertical density structure.  

As for the main differences between models and observations, the synthetic models overpredict the CO peak brightness temperature beyond $\sim$100 au by about 30-50\%. 
Since the center of the CO line in those regions is optically thick, this discrepancy suggests that our model overpredicts the temperature of the CO emitting layer, which is typically located a few scale heights above the disk midplane \citep{ Weaver+2018,Law+2022,Law+2023,Galloway_exoALMA}. 
In our model, the temperature is calculated using RADMC3D based on the stellar luminosity and disk structure.
Reducing the temperature of the outer disk by 40\% would require either lowering the stellar luminosity from about 1 $L_\odot$ to about 0.3 $L_\odot$, which is inconsistent with observations of LkCa~15, altering the opacity of the small grains which intercept the stellar radiation, or reducing the flaring angle of the outer disk \footnote{In first approximation, the disk temperature scales as $(\alpha L_\star)^{1/4}$, where $\alpha$ is the incident angle between the stellar radiation and the disk surface \citep{Dullemond+2001}}. 
Computing the gas temperature and abundance self-consistently with a physical-chemical model may also lower the temperature in the outer disk.
Finally, the zero velocity channel of the data reveals an s-shaped kink at about $\sim$20 au from the center of the disk which shows up as two red spots in the residual channel maps. This kink is along the minor axis of the disk like the bridge of emission identified in the continuum, but lies closer radially to the peak of the first continuum ring.
These kinks can be interpreted as the effect of kinematic perturbations caused by an embedded planet, though their origin is still under debate \citep{Pinte+2023}.
We refer the reader to \citet{Pinte_exoALMA}, for a discussion of the velocity kinks in the exoALMA sample.

\section{Discussion and Conclusion} 
\label{sec:discuss}


As discussed in the Introduction, it has long been speculated that LkCa~15's disk hosts accreting protoplanets. 
The presence of the large continuum cavity was the first motivation for these claims, followed by optical and NIR observations of candidate planets within the cavity. 
High angular resolution observations from ALMA showing bright continuum rings also supported these theories with speculation that they could be due to dust trapping from planet-driven pressure bumps. 
Yet, the proposed protoplanets have been disputed with scattered light imaging suggesting that the compact emission observed in the NIR could instead arise from diffused dust particles with the dust cavity. 
This hypothesis is consistent with the faint 0.87~mm dust continuum and CO emission recorded within the innermost dust ring, discussed in Section~\ref{sec:obs}.
Indeed, the (sub)micron-sized particles responsible for the observed scattered light emission should be coupled to the gas, and, therefore, follow a radial distribution similar to that of CO molecules. 

The simultaneous analysis of high angular resolution continuum and CO observations with hydro and radiative transfer models allow for a detailed study of the cavity identified in the continuum. 
In Section \ref{subsec:model:cont}, we showed that the dust continuum emission is consistent with dynamical perturbations from three planets orbiting at 29, 51, and 80 au, with masses of 3-5 $M_\textrm{J}$, 0.4~$M_\textrm{J}$, and 0.25~$M_\textrm{J}$, respectively. The three rings are likely circular and coplanar (based on the same center and inclination) suggesting that the putative planets would be on circular orbits. In that context, a single but massive planet is required to carve the large inner cavity. 
However, in Section~\ref{subsec:model:CO}, we found that the presence of a massive innermost planet is inconsistent with observed CO emission, which would instead require a less massive planet ($M_p \lesssim 1 $ $M_\textrm{J}$) even in the most conservative case in which CO photodissociation is not considered. In practice, there is a delicate balance in the mass of the innermost planet.
The planet's mass cannot be so low that it does not clear out a deep cavity in the continuum, but it cannot be so massive that it leaves a signature in the CO emission.
Our results indicate that such a delicate interplay cannot be obtained with a single planet.

We conclude this section by discussing how our results might depend on the initial choice of model parameters kept fixed in our simulations. In particular, the relative distribution between gas and dust depends strongly on the aerodynamic coupling between dust and gas, which is controlled by the ratio between the dust Stokes number St and viscosity parameter $\alpha$ \citep{Dipierro+2016}. The Stokes number depends on the dust size $a$ and shape \citep{Pinte+2019}, which are fixed, and the gas surface density $\Sigma_{g}$, which depends on the choice of $\alpha$. 
Our models indicate that for the given choice of $a$ and $\alpha$, the dust is too coupled to the gas resulting in the large cavity, or the lack thereof, being visible in both tracers. 
To reduce the coupling, we could increase St/$\alpha$ by increasing the dust size or decreasing $\alpha$. A lower $\alpha$ would mean that massive planets would clear more gas within the dust cavity and excite instabilities, such as the Rossby Wave Instability \citep{Lovelace+1999,Li+2000}, resulting in more asymmetric substructures. Therefore, the inner planet mass would have to be lowered even more to match the observations. 

Our model assumes that grains with size $a=0.14$~mm are the main carriers of the 0.87~mm dust opacity. 
This is a good assumption in the case of a standard grain size distribution ($n(a)\propto a^{-3.5}$), but it might be inappropriate if the grain size distribution is strongly skewed toward mm/cm size grains. In this case, a larger dust size would result in a larger Stokes number and less coupling. A less massive inner planet may still be able to trap these larger dust grains while less effectively clearing a cavity in gas, therefore bringing us closer towards the differences in gas and dust that we observe.
Constraining the grain size distribution in LkCa 15's disk, however, requires high angular resolution observations across a broad range of wavelengths (e.g., 0.8-3mm), which, as of the time of this study, do not exist \citep[see, e.g.][]{Guidi+2022}). 

Another alternative to reduce the coupling is to reduce $\Sigma_{g}$. Doing this while holding $\alpha$ and $a$ fixed, however, would make the cavity more clear of gas which is the opposite of what we would want. If a planet exists in the cavity, then, its mass being low seems to be a robust result even with changes to the coupling.

Alternatively, a chain of properly spaced low-mass planets orbiting within 30 au from the central star could also explain the observed dust and gas morphology, as suggested by \citet{Leemker+2022}.
If this is the case, observations of the dust and CO emission at higher angular resolution and sensitivity might reveal rings and gaps within the dust cavity created by these planets. 
Alternatively, physical processes that do not require the presence of planets could be responsible for LkCa~15's dust cavity.  
These are discussed in other exoALMA papers and we refer the reader to those for more details \citep{Barraza_exoALMA}. 
In any case, high-angular resolution observations of gas tracers more optically thin than $^{12}$CO, such as $^{13}$CO and C$^{18}$O, would help to better constrain the gas density and CO photodissociation within the dust cavity and inform about its origin. 
This goal could be achieved by combining exoALMA observations of these CO isotopologues, which achieved an angular resolution between 0.1\arcsec--0.3\arcsec \citep{Teague_exoALMA}, with future ALMA long baseline observations. As of now, the exoALMA $^{13}$CO resolution only allows for $\sim$1-2 beams across the disk cavity meaning that structures will only be marginally resolved at best.

In conclusion, in this work, we have shown that observing molecular line emission at the same angular resolution of the sharpest dust continuum images is a key step in investigating the origin of dust substructures.  
In the case of LkCa~15, observations of the CO line emission that achieve a spatial resolution of about 6~au reveal the presence of gas inside a dust-depleted cavity. 
Under reasonable assumptions on the disk's physical properties, we demonstrated that these observations are inconsistent with the postulated presence of massive planets orbiting between 10-30 au. 
We show that planet-disk interaction models reproduce the overall radial and azimuthal morphology of dust and gas emissions well. 
However, the estimated planet masses are less than the mass of Jupiter, and, for this reason, might be too faint to be directly detected by existing NIR high-contrast cameras.

\section*{Acknowledgments}

We thank the anonymous referee for their support in refining this paper. This paper makes use of the following ALMA data: ADS/JAO.ALMA\#2021.1.01123.L. ALMA is a partnership of ESO (representing its member states), NSF (USA) and NINS (Japan), together with NRC (Canada), MOST and ASIAA (Taiwan), and KASI (Republic of Korea), in cooperation with the Republic of Chile. The Joint ALMA Observatory is operated by ESO, AUI/NRAO and NAOJ. The National Radio Astronomy Observatory is a facility of the National Science Foundation operated under cooperative agreement by Associated Universities, Inc. We thank the North American ALMA Science Center (NAASC) for their generous support including providing computing facilities and financial support for student attendance at workshops and publications.

AI and CHG acknowledge support from the National Aeronautics and Space Administration under grant No. 80NSSC18K0828.
CHG gratefully acknowledges the support from a LANL/CSES Student Fellow project. CHG, HL, and SL also gratefully acknowledge the support from LANL/LDRD Project 20240039DR. This research used resources provided by the Los Alamos National Laboratory Institutional Computing Program, which is supported by the U.S. Department of Energy National Nuclear Security Administration under Contract No. 89233218CNA000001.

JB acknowledges support from NASA XRP grant No. 80NSSC23K1312. MB has received funding from the European Research Council (ERC) under the European Union’s Horizon 2020 research and innovation programme (PROTOPLANETS, grant agreement No. 101002188).
PC acknowledges support by the Italian Ministero dell'Istruzione, Universit\`a e Ricerca through the grant Progetti Premiali 2012 – iALMA (CUP C52I13000140001) and by the ANID BASAL project FB210003.
SF is funded by the European Union (ERC, UNVEIL, 101076613). Views and opinions expressed are however those of the author(s) only and do not necessarily reflect those of the European Union or the European Research Council. Neither the European Union nor the granting authority can be held responsible for them. SF acknowledges financial contribution from PRIN-MUR 2022YP5ACE.
DF has received funding from the European Research Council (ERC) under the European Union’s Horizon 2020 research and innovation programme (PROTOPLANETS, grant agreement No. 101002188).
MF is supported by a Grant-in-Aid from the Japan Society for the Promotion of Science (KAKENHI: No. JP22H01274).
CH acknowledges support from NSF AAG grant No. 2407679.
JDI acknowledges support from an STFC Ernest Rutherford Fellowship (ST/W004119/1) and a University Academic Fellowship from the University of Leeds.
Support for AFI was provided by NASA through the NASA Hubble Fellowship grant No. HST-HF2-51532.001-A awarded by the Space Telescope Science Institute, which is operated by the Association of Universities for Research in Astronomy, Inc., for NASA, under contract NAS5-26555.
CL has received funding from the European Union's Horizon 2020 research and innovation program under the Marie Sklodowska-Curie grant agreement No. 823823 (DUSTBUSTERS) and by the UK Science and Technology research Council (STFC) via the consolidated grant ST/W000997/1.
FMe received funding from the European Research Council (ERC) under the European Union's Horizon Europe research and innovation program (grant agreement No. 101053020, project Dust2Planets).
CP acknowledges Australian Research Council funding  via FT170100040, DP18010423, DP220103767, and DP240103290.
DP acknowledges Australian Research Council funding via DP18010423, DP220103767, and DP240103290.
GR acknowledges funding from the Fondazione Cariplo, grant no. 2022-1217, and the European Research Council (ERC) under the European Union’s Horizon Europe Research \& Innovation Programme under grant agreement no. 101039651 (DiscEvol). Views and opinions expressed are however those of the author(s) only, and do not necessarily reflect those of the European Union or the European Research Council Executive Agency. Neither the European Union nor the granting authority can be held responsible for them.
JS has received funding from the European Research Council (ERC) under the European Union’s Horizon 2020 research and innovation programme (PROTOPLANETS, grant agreement No. 101002188). Computations have been done on the ’Mesocentre SIGAMM’ machine hosted by Observatoire de la Côte d’Azur. GWF acknowledges support from the European Research Council (ERC) under the European Union Horizon 2020 research and innovation program (Grant agreement no. 815559 (MHDiscs)).
GWF was granted access to the HPC resources of IDRIS under the allocation A0120402231 made by GENCI.
AJW has received funding from the European Union’s Horizon 2020
research and innovation programme under the Marie Skłodowska-Curie grant agreement No 101104656. 
H-WY\ acknowledges support from National Science and Technology Council (NSTC) in Taiwan through grant NSTC 113-2112-M-001-035- and from the Academia Sinica Career Development Award (AS-CDA-111-M03).
TCY acknowledges support by Grant-in-Aid for JSPS Fellows JP23KJ1008.
Support for BZ was provided by The Brinson Foundation.
We acknowledge the use of ChatGPT (OpenAI) for minor refinements to the language of the text.

\appendix

\section{Dust appearance vs Robust parameter}
\label{sec:rob}

\begin{figure*}
    \centering
\includegraphics[width=\linewidth]{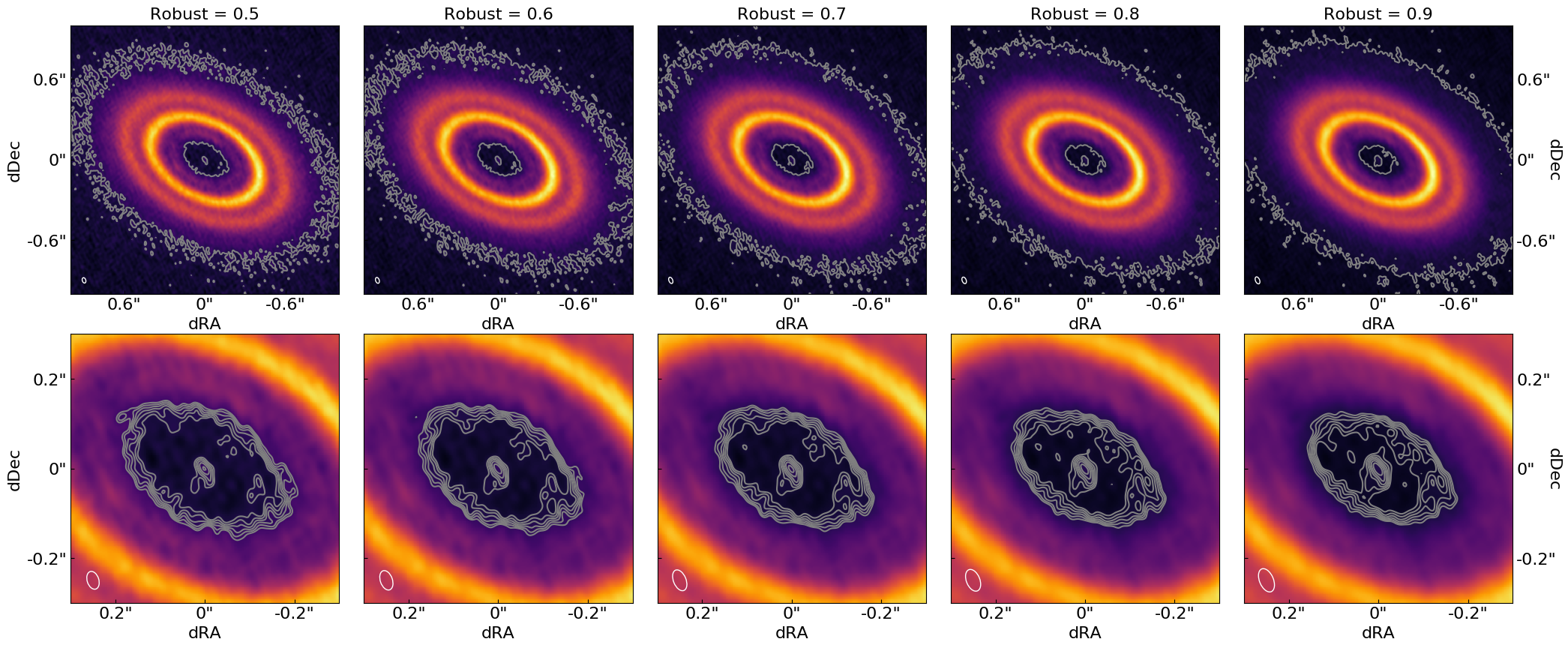}
    \caption{A figure showing the effects of using different robust parameters when imaging the continuum data. (Top Row) An image of the full disk with contours of 4 times the rms noise of each image. As the robust parameter increases, the compact source slowly becomes part of more extended dust emission present in the cavity. (Bottom Row) The inner cavity with contours of 3, 4, 5, 6, 7, and 8 times the rms noise. As the robust parameter increases, emission near the central source begins to connect with emission near the inner continuum ring. More material also becomes visible in the cavity at lower rms with the higher robust parameter.}
    \label{fig:Robust}
\end{figure*}

Figure~\ref{fig:Robust} shows images of the dust continuum emission obtained using robust parameters between 0.5 and 0.9, resulting in synthesized beam sizes increasing from $0.041"\times0.026"$ to $0.055"\times0.031"$, and rms noises decreasing from 11.7 $\mu Jy$ $\text{beam}^{-1}$ to 10.7 $\mu Jy$  $\text{beam}^{-1}$. 

\section{Disk geometry and properties of the dust rings}
\label{sec:ring_geo}
To measure the properties of the dust rings, we start by fitting ellipses to the continuum image following the procedure from \citet{Huang+2018}. 
This approach uses the Python-based Markov Chain Monte-Carlo sampler \textit{emcee} \citep{Foreman-Mackey+2013} to find the best-fitting ellipse for intensity maxima or minima within a given radial interval.
All the rings and gaps have similar centers, inclinations, and position angles, indicating that these substructures are concentric and coplanar. 
Using the average value of the ring centers and orientation, we then deproject the continuum emission and create the azimuthally averaged profile shown in the left panel of Figure~\ref{fig:moment} and all panels of Figure~\ref{fig:mod_cont_rad}. 
This profile clearly shows the presence of central emission and of three rings with radii of about 43 au, 69 au, and 100 au. 

Next, we measure the continuum rings' width in the image and visibility plane. Both the image and visibility fitting require an additional diffused component to account for the extended emission observed beyond 150 au, which here is modeled as an additional broader ring.   
In the case of the image plane, we use \textit{emcee} to fit 1D Gaussian functions to the azimuthally averaged continuum intensity profile shown in the left panel of Figure~\ref{fig:moment}.
We then take the width of each component and deconvolve it via a simple sum of squares calculation using the geometrical mean of the synthesized beam size. 
We also deconvolve the amplitudes of the Gaussians using a ratio of the measured $\sigma$ to the deconvolved $\sigma$. 
This provides the widths listed in column 9 of the image fitting section of Table \ref{tab:rings}. 
For the analysis in the visibility plane, we use the \textit{Galario} software package \citep{Tazzari+2018}. Galario takes an input model image, which in this case consists of the sum of Gaussian rings, and calculates the corresponding visibilities. Using \textit{emcee}, we then find best-fitting model parameters by comparing observed and theoretical visibilities. In both cases, we only fit for one center, inclination, and position angle, assuming that the rings are concentric and coplanar. 
The image and visibility fittings deliver consistent results regarding the disk orientation and the ring properties. We find that the ratios between the width of the rings (defined as the dispersion of the Gaussian function) and their radius are 0.19, 0.11, and 0.14 for the innermost, middle, and outermost rings, respectively. 
The implication of the widths of the rings in terms of dust-trapping models is discussed further in \citet{Stadler_exoALMA}.

\begin{deluxetable*}{cccccccccc}[!htb]
\tablecaption{\label{tab:table_appen}}
\tablecolumns{10}
\tablenum{2}
\tablewidth{0pt}
\tablehead{
\colhead{Fit} &
\colhead{Feature} &
\colhead{$\Delta$x} &
\colhead{$\Delta$y} &
\colhead{r$_{0}$} &
\colhead{r$_{0}$} &
\colhead{Incl.} &
\colhead{P.A.} &
\colhead{$\sigma$} &
\colhead{Amplitude} \\ &
\colhead{} &
\colhead{(mas)} &
\colhead{(mas)} &
\colhead{(mas)} &
\colhead{(au)} &
\colhead{(degrees)} &
\colhead{(degrees)} &
\colhead{(au)} &
\colhead{(Jy/arcsec$^{2}$)}\\
\colhead{(1)}&
\colhead{(2)}&
\colhead{(3)}&
\colhead{(4)}&
\colhead{(5)}&
\colhead{(6)}&
\colhead{(7)}&
\colhead{(8)}&
\colhead{(9)}&
\colhead{(10)}
}
\startdata
Ellipse & B43 & -37.1$^{+2.4}_{-2.3}$& -76.9$\pm2.3$ & 267.6$^{+2.9}_{-3.0}$ & 42.5$\pm0.5$ & 49.1$\pm1.1$ & 63.0$^{+1.4}_{-1.5}$ & - & -  \\
Fitting & D46 & -34.2$^{+3.3}_{-3.0}$ & -77.2$\pm3.0$ & 290.6$^{+3.9}_{-3.6}$ & 46.2$\pm0.6$ & 48.1$^{+1.8}_{-1.6}$ & 66.0$^{+1.7}_{-1.8}$ & - & - \\
 & B69 & -33.7$\pm0.7$ & -78.8$\pm0.6$ & 434.2$\pm0.8$ & 69.0$\pm0.1$ & 50.5$\pm0.2$& 61.7$\pm0.2$ & - & -  \\
 & D87 & -40.1$^{+1.3}_{-1.2}$ & -81.1$\pm1.1$& 547.5$^{+1.6}_{-1.5}$ &87.0$^{+0.3}_{-0.2}$ & 50.3$^{+0.2}_{-0.3}$& 62.0$\pm0.3$ & - & -  \\
& B101 &  -37.6$^{+1.4}_{-1.3}$ & -81.9$\pm1.1$ & 634.4$^{+1.7}_{-1.6}$ & 100.8$\pm0.3$ & 51.0$\pm0.2$ & 61.7$\pm0.3$& - & - \\
& Average & -36.5$^{+1.8}_{-1.7}$ & -79.2$\pm1.6$ & - & - & 49.8$\pm0.7$ & 62.9$\pm0.8$ & - & - \\
 \hline
 Image  & Central Source & -36.5$^{+1.8}_{-1.7}$ & -79.2$\pm1.6$ & - & - & [49.8] & [62.9] & 2.5$\pm1.8$ & 0.064$^{+4.022}_{-0.047}$ \\
 Fitting & B44 & -36.5$^{+1.8}_{-1.7}$ & -79.2$\pm1.6$ & 277.9$^{+1.2}_{-1.1}$ & 44.2$\pm0.2$ & [49.8] & [62.9] & 8.5 $\pm0.2$ & 0.111$\pm0.008$ \\
 & B69 & -36.5$^{+1.8}_{-1.7}$ & -79.2$\pm1.6$ & 436.3$^{+0.3}_{-0.2}$ & 69.3$\pm0.04$ & [49.8] & [62.9] & 8.0$\pm0.05$ & 0.451$\pm0.007$ \\
 & B100 & -36.5$^{+1.8}_{-1.7}$ & -79.2$\pm1.6$ & 631.0$\pm 0.6$ & 100.3$\pm0.1$ & [49.8] & [62.9] & 14.5$\pm0.1$ & 0.232$\pm0.006$ \\
 & B107 & -36.5$^{+1.8}_{-1.7}$ & -79.2$\pm1.6$ & 673.6$^{+7.2}_{-6.9}$ & 107.0$^{+1.2}_{-1.1}$ & [49.8] & [62.9] & 51.7$\pm0.5$ & 0.073$^{+0.003}_{-0.002}$ \\
 \hline
 Visibility  & Central Source & -40.4$\pm 0.3$ & -84.0$\pm 0.2$ & - & - & 50.7$\pm0.1$ & 61.7 $\pm0.1$ & 1.47$^{+0.7}_{-0.6}$ & 0.287$^{+0.486}_{-0.143}$ \\
Fitting & B43 & -40.4$\pm 0.3$ & -84.0$\pm 0.2$ & 272.4$^{+1.7}_{-1.6}$ & 43.3$\pm$0.3 & 50.7$\pm0.1$ & 61.7 $\pm0.1$ & 8.4$\pm0.4$ & 0.115$\pm0.003$ \\
 & B69 & -40.4$\pm 0.3$ & -84.0$\pm 0.2$ & 432.7$\pm0.4$ & 68.8$\pm 0.1$ & 50.7$\pm0.1$ & 61.7 $\pm0.1$ & 7.67$\pm0.1$ & 0.478$\pm0.003$\\
 & B100 & -40.4$\pm 0.3$ & -84.0$\pm 0.2$ & 629.8$^{+0.8}_{-0.9}$& 100.1$\pm 0.1$ & 50.7$\pm0.1$ & 61.7 $\pm0.1$ & 14.1$\pm0.2$ & 0.235$\pm0.002$ \\
 & B104 & -40.4$\pm 0.3$ & -84.0$\pm 0.2$ & 654.8$^{+9.6}_{-9.3}$ & 104.1$\pm1.5$ & 50.7$\pm0.1$ & 61.7 $\pm0.1$ & 48.0$^{+0.7}_{-0.8}$ & 0.079 $\pm0.002$ \\
\enddata
\vspace{0.5cm}
\caption{Results of various fits to the rings of LkCa~15. The table is divided according to the three different fitting procedures outlined in the text, the ellipse fitting routine of \citet{Huang+2018}, image fitting of the observed radial profile, and fitting in the visibility space using \textit{galario}. The ellipse fitting finds both rings (B) and gaps (D). Only the ellipse fitting fits for many concentric and coplanar rings leading to different values for each ring in columns (3), (4), (7), and (8). We average these values together and use them to deproject the disk and create the radial profile used in the image fitting. The ellipse fitting, however, does not account for the width or amplitude of the rings. In columns (9) and (10) we list the deconvolved standard deviations and amplitudes from our image fitting and visibility fitting. Since \textit{galario} works in the visibility space, the values were already deconvolved.  In the image domain  we use a simple sum of squares formula to deconvolve the beam and get the standard deviation. This deconvolved standard deviation is then used to deconvolve the amplitude.} 
\label{tab:rings}
\end{deluxetable*}

\section{CO channel maps}
\label{sec:co_channel}
\begin{figure*}
    \centering
    \includegraphics[width=\linewidth]{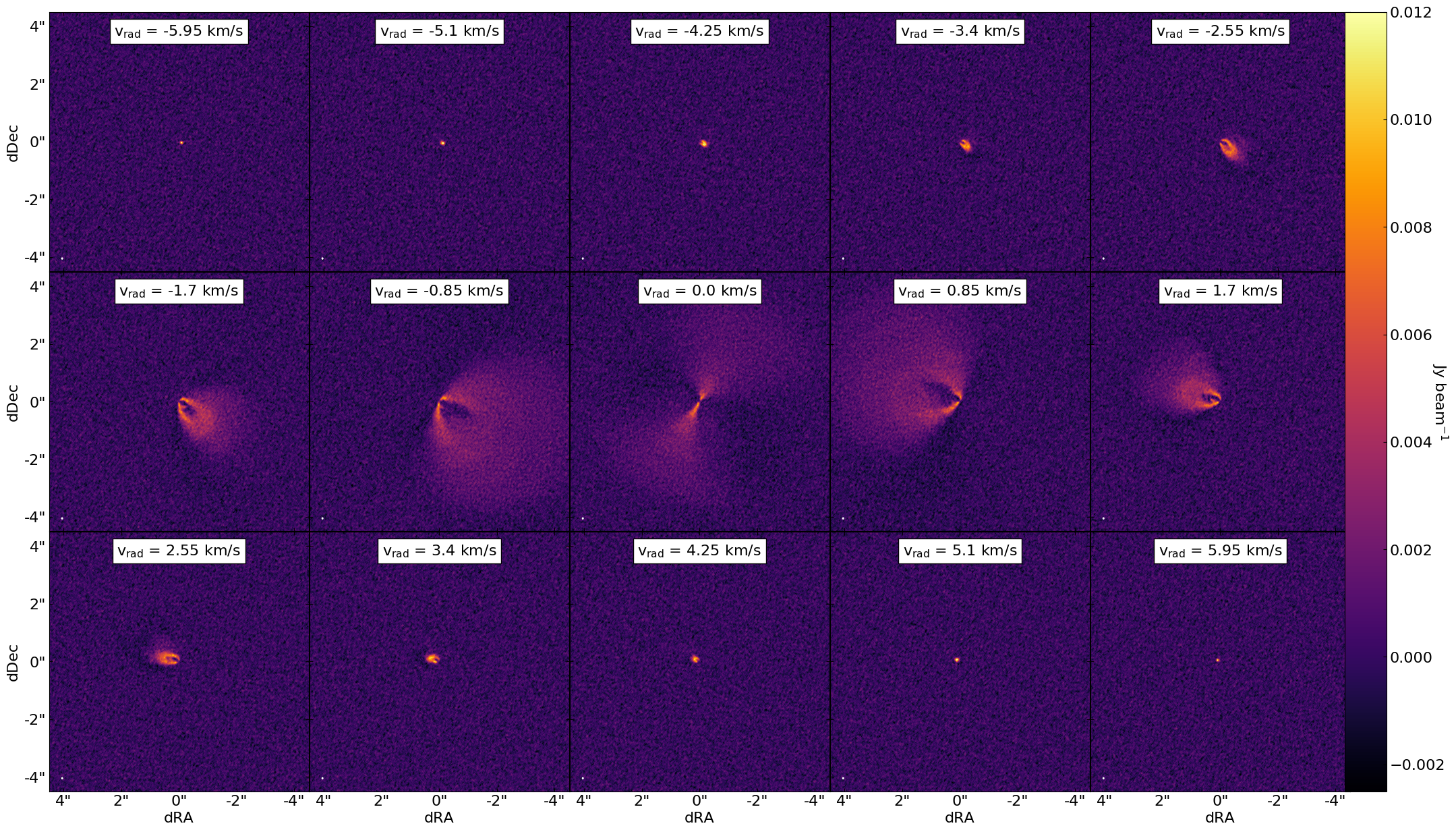}
    \caption{Channel maps of the combined $^{12}$CO data cube. The velocity resolution is quite poor at only 0.85 km/s as the original observations were focused on maximizing continuum sensitivity and resolution, leaving CO to be detected serendipitously. Whereas this work focuses mostly on the interior of the disk, these maps show its full extent out to $\sim$600 AU. We use v$_{lsr}$ = 6.3 km/s to calculate v$_{rad}$.}
    \label{fig:channel}
\end{figure*}

In Figure~\ref{fig:channel}, we show the channel maps of the exoALMA + cycle 6 full data cube. In other figures we limited the extent of the images to the inner $\sim$100 au whereas here we show the full extent of the disk out to $\sim$600 au. The velocity resolution is only 0.85 km/s as the exoALMA data were spectrally averaged in order to combine with the cycle 6 data. Combining these data helped to reduce the effects of spatial filtering, a result of the cycle 6 data not having shorter baselines in the uv coverage. This lead to large regions of negative emission in the cycle 6 channel maps.

\section{Disk-Planet Models}
In this section we provide more details about the simulations we run using LA-COMPASS and RADMC. Many recent studies using hydro simulations of planets embedded in protoplanetary disks show the importance of three dimensions in capturing all of the appropriate accretion flows onto the planet. Namely, studies like those of \citet{Szulagyi+2016} showed that accretion flows to the planet from the surrounding cicumstellar disk originate from mid to high latitudes above the disk midplane and then flow outwards through the CPD midplane. This meridional cycling of material is a behavior that can only be captured in 3D, however, at the highest of latitudes above the midplane of the CPD, material accretes at a significant fraction of the free-fall velocity. In order to satisfy the CFL criteria with these rapid motions, the time step must be very small. This makes 3D simulations very computationally expensive to run, even if only run for $\sim10^{2}$ orbits. As we are interested in the evolution of the circumstellar disk on timescales of $10^{3}$ orbits in this study, 3D simulations are computationally unfeasible. 

To further reduce computational costs we ran the simulations with only one grain size. Our choice of grain size arises from taking the integral average of the opacity, weighted by the MRN distribution \citep{MRN+1977}. When this is done, one finds that the grain size that tends to dominate the opacity is $a\sim\frac{\lambda}{2\pi}$ which for our observations corresponds to 0.14 mm grains.

\begin{figure*}
    \centering
    \includegraphics[width=\linewidth]{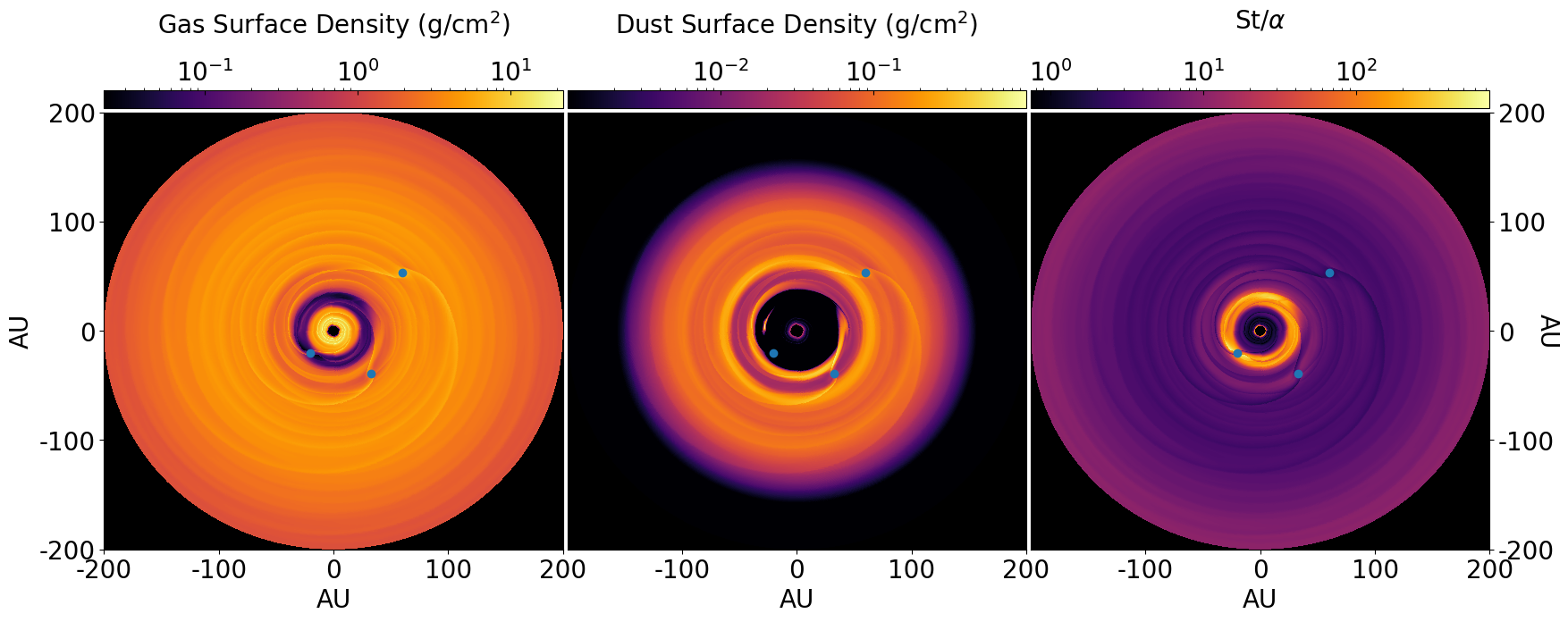}
    \caption{A snapshot of our 2D hydro runs from LA-COMPASS. Each snapshot is taken at 5000 orbits. The blue dots represent the location of the planets in the simulation. Left: $\Sigma_{g}$ of the 3.8 M$_{\textrm{J}}$ model. The massive planet in the inner disk clears out a cavity which later becomes visible in our synthetic images generated from RADMC3D. Middle: $\Sigma_{d}$ of the 3.8 M$_{\textrm{J}}$ model. This configuration of planets is able to clear out a cavity in the dust and reproduce three rings, but the rings are more asymmetric than those seen in the observations Right: St/$\alpha$ of the 3.8 M$_{\textrm{J}}$ model. This image is anti-correlated with the gas surface density as St $\propto\Sigma_{g}^{-1}$ and other quantities that set St are held constant in time.}
    \label{fig:sigma_gas}
\end{figure*}

Figure \ref{fig:sigma_gas} shows the gas and dust surface densities as well as Stokes/$\alpha$ for the $\sim4M_{\textrm{J}}$ model shown in the right panel of Figure \ref{fig:mod_cont_im} (taken at 5000 orbits for the innermost planet). The massive planet clears out a cavity in the gas which is visible in the synthetic images shown in Figure \ref{fig:mod_temp_im}. Much of the dust has already drifted inwards, leading to the smaller radial extent of material in the dust surface density. The larger cavity in dust shows the extent of the decoupling of the dust grains. While the decoupling can typically be quantified with the ratio of Stokes/$\alpha$, here we are keeping the grain size and $\alpha$ constant. Thus, our plot of St/$
\alpha$ is controlled by the gas surface density and the two are anti-correlated. 

To perform the temperature calculations using RADMC3D we initially degrade the resolution of our grid from LA-COMPASS to decrease the computation time of our thermal monte carlo simulations. Our RADMC grid is $205\times100\times80$ ($N_{r}\times N_{\phi}\times N_{\theta}$), covering the same radial and azimuthal extent of the LA-COMPASS simulations. We linearly interpolate from the LA-COMPASS to RADMC grids, taking care to ensure that the interpolation does not change the gas or dust mass by more than $\sim10\%$. In the polar angle we run from 60$^{\circ}$ to 90$^{\circ}$ (midplane) and assume that the disk is symmetric about the midplane. RADMC3D also allows for a variable number of photons to be used in the thermal simulation, however, too few photons results in temperature profiles that are dominated by monte carlo noise. We tried running the thermal calculation with up to $2\times10^{9}$ photons, but found that $2.5\times10^{8}$ photons achieved a good balance between computational time and reducing the noise. 

\begin{figure}
    \centering
    \includegraphics[width=\linewidth]{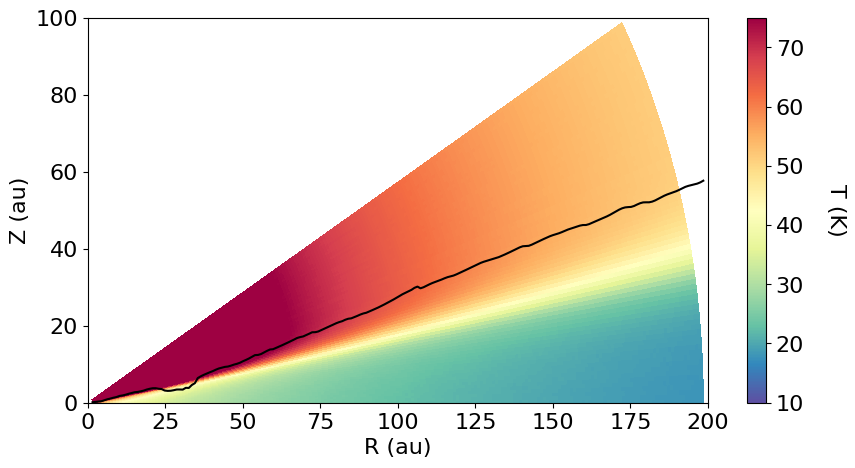}
    \caption{Color plot of an azimuthal slice of the 3D temperature calculated with RADMC3D. We also show the photodissociation surface calculated in Section \ref{subsec:setup} as a black line. There is a dip in the photodissociation surface associated with the massive planet at 25 au in this model.}
    \label{fig:z_phot_temp}
\end{figure}

Figure \ref{fig:z_phot_temp} shows an azimuthal slice of the temperature calculated with RADMC3D. The disk naturally settles into two layers (midplane and surface). Also overplotted is the photodissociation surface calculated as described in Section \ref{subsec:setup}. The drop in the photodissociation surface at $\sim25-30$ au comes from the massive planet removing much of the gas at that location. 

\section{Photodissociation}
\label{sec:photodis}
The prescription for photodissociation that we take from \citet{Rosenfeld+2013} and \citet{Qi+2011} does not fully capture the complexity of this physical process.
The CO photodissociation threshold depends on the amount of shielding from dust and H$_{2}$, the CO self-shielding, and the shape of the radiation field. The surface density threshold of $5\times10^{20}$ cm$^{-2}$ corresponds to an extinction of A$_{V}$ = 1 when assuming a dust-to-gas ratio of 1\% \citep{Qi+2011}. 
However, if the dust-to-gas ratio is reduced, the CO photodissociation will occur deeper in the disk atmosphere due to a reduction in the shielding from dust. In \citet{Qi+2011}, the threshold value is reached above the $\tau = 1$ surface for the dust, implying that a sufficient column for shielding has already been reached in H$_{2}$ and CO. However, adjustments in the assumed CO abundance would either lessen or magnify the effects of CO self-shielding and again shift the depth at which the photodissociation surface forms. To complicate matters further, the shielding also depends on the gas temperature \citep{Visser+2009}. We assumed that the gas temperature was the same as the dust temperature in order to decouple the radiative transfer calculations from the hydrodynamic evolution. This allowed for a more thorough exploration of the model parameter space, but physical-chemical modeling has shown that this is not always true above the disk midplane  \citep{Bruderer+2012,Bruderer+2013}.

\begin{figure*}
    \centering
    \includegraphics[width=\linewidth]{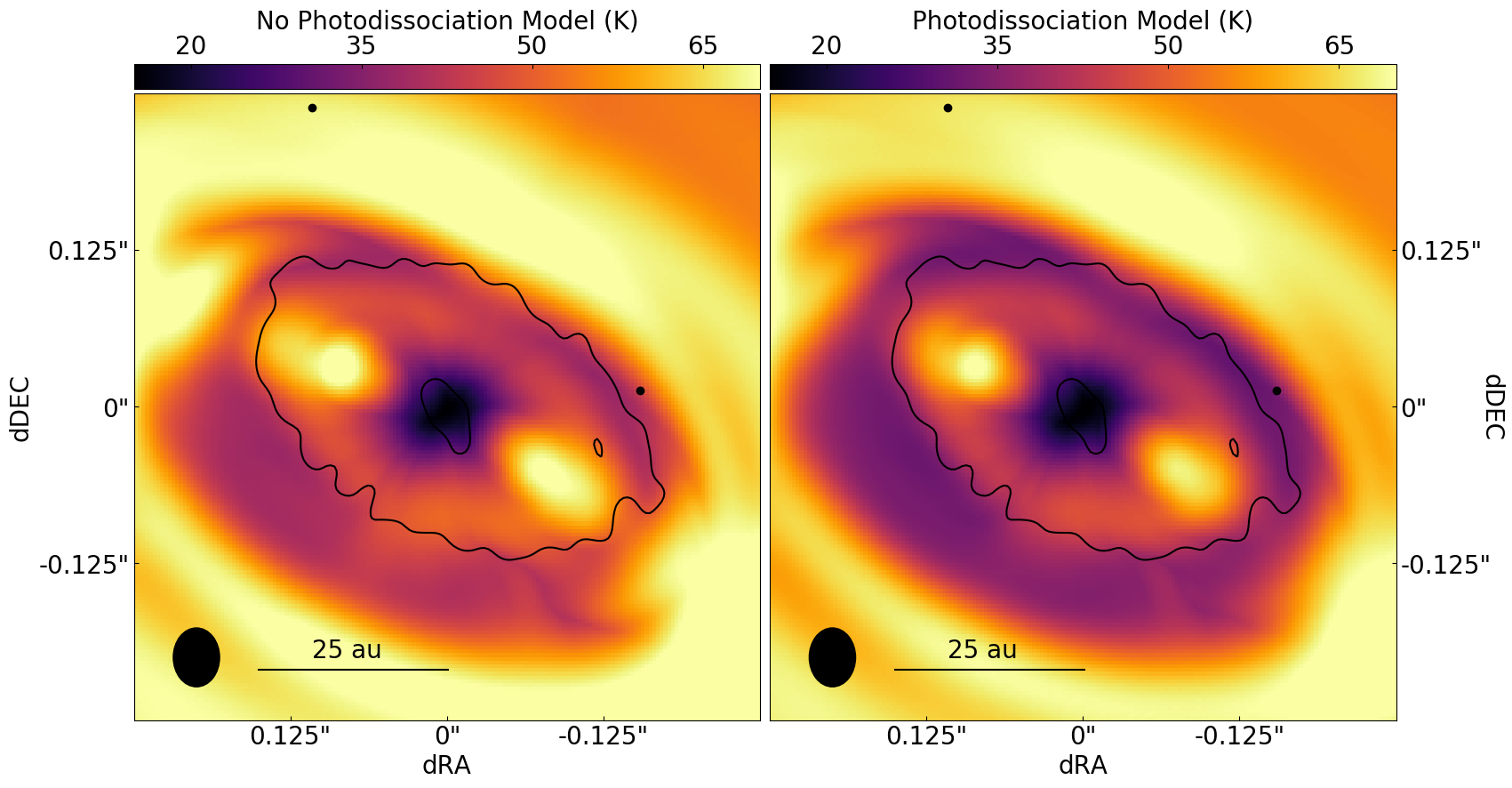}
    \caption{Left: The same image shown in the right panel of Figure \ref{fig:mod_temp_im} zoomed in to the inner $0.5\arcsec\times0.5\arcsec$ with no photodissociation. Right: The same as Left, but including our prescription for photodissociation. There are only small changes in the brightness of the emission in the cavity between the two images. Because of this, we can reasonably place an upper limit on the mass of any possible planets. As in Figure~\ref{fig:mod_cont_im}, we show the 5$\sigma$ contour of the continuum to emphasize the location of the dust cavity and mark the locations of the planets with black dots.}
    \label{fig:mod_temp_photodissociation}
\end{figure*}

Ultimately, in the absence of physical-chemical modeling, full radiative-hydrodynamic calculations, or chemical networks like those of \citet{vantHoff+2017}, a number of assumptions had to be made about gas and dust abundances and temperatures. 
In the future, a more thorough treatment of the radiative and thermal physics, or the use of the aforementioned chemical networks, will preclude the assumptions made in this work. In that case, the implementation of effects like CO photodissociation can be done more accurately. Regardless, Figure~\ref{fig:mod_temp_photodissociation} shows the effect of both excluding and including photodissociation in our CO models. Since the differences in the two images are marginal, we have confidence that the results derived on planet masses in this work are robust.

\clearpage


\bibliographystyle{aasjournal}

\end{document}